\documentclass[10pt,journal, twocolumn, final]{IEEEtran}
\usepackage[T1]{fontenc}
\usepackage[latin9]{inputenc}
\usepackage{verbatim}
\usepackage{mathrsfs}
\usepackage{amsmath}
\usepackage{amsthm}
\usepackage{amssymb}
\usepackage{graphicx}
\usepackage{esint}

\makeatletter
\theoremstyle{definition}
\newtheorem{assumption}{Assumption}
  \theoremstyle{plain}
  \newtheorem{thm}{\protect\theoremname}

\usepackage{cite}
\allowdisplaybreaks
\makeatother

\providecommand{\theoremname}{Theorem}

\begin{document}

\title{Model-based reinforcement learning in differential graphical games\thanks{Rushikesh Kamalapurkar, is with the School of Mechanical and Aerospace
Engineering, Oklahoma State University, Stillwater, OK, USA. Email:
rushikesh.kamalapurkar@okstate.edu. Justin R. Klotz, Patrick Walters,
and Warren E. Dixon are with the Department of Mechanical and Aerospace
Engineering, University of Florida, Gainesville, FL, USA. Email: \{jklotz,
walters8, wdixon\}@ufl.edu.}\thanks{This research is supported in part by National Science Foundation
award numbers 1217908 and 1509516, and Office of Naval Research award
numbers N00014-13-1-0151 and N00014-16-1-2091. Any opinions, findings
and conclusions or recommendations expressed in this material are
those of the authors and do not necessarily reflect the views of the
sponsoring agency.}}

\author{Rushikesh Kamalapurkar, Justin R. Klotz, Patrick Walters, and Warren
E. Dixon}
\maketitle
\begin{abstract}
This paper seeks to combine differential game theory with the actor-critic-identifier
architecture to determine forward-in-time, approximate optimal controllers
for formation tracking in multi-agent systems, where the agents have
uncertain heterogeneous nonlinear dynamics. A continuous control strategy
is proposed, using communication feedback from extended neighbors
on a communication topology that has a spanning tree. A model-based
reinforcement learning technique is developed to cooperatively control
a group of agents to track a trajectory in a desired formation. Simulation
results are presented to demonstrate the performance of the developed
technique.
\end{abstract}

\section{Introduction}

In the past few decades, reinforcement learning (RL)-based techniques
have been established as primary tools for online real-time optimization
\cite{Sutton1998,Bertsekas2007,Vamvoudakis2010,Modares.Lewis.ea2014,Konda2004,Zhang.Cui.ea2013,Heydari.Balakrishnan2013}.
RL techniques are valuable not only for optimization but also for
control synthesis in complex systems such as a distributed network
of cognitive agents. Combined efforts from multiple autonomous agents
can yield tactical advantages including: improved munitions effects;
distributed sensing, detection, and threat response; and distributed
communication pipelines \cite{Ren.Beard2008,Semsar-Kazerooni.Khorasani2013}.
While coordinating behaviors among autonomous agents is a challenging
problem that has received mainstream focus, unique challenges arise
when seeking optimal autonomous collaborative behaviors. For example,
most collaborative control literature focuses on centralized approaches
that require all nodes to continuously communicate with a central
agent, yielding a heavy communication demand that is subject to failure
due to delays, and missing information\cite{Murray2007}. Furthermore,
the central agent is required to carry enough on-board computational
resources to process the data and to generate command signals. These
challenges motivate the need to minimize communication for guidance,
navigation and control tasks, and to distribute the computational
burden among the agents.

Since all the agents in a network have independent collaborative or
competitive objectives, the resulting optimization problem is a multi-objective
optimization problem. Differential game theory is often used to define
optimality in multi-objective optimization problems \cite{Tidball.Altm1996,Tidball.Pourtallier.ea1997,Isaacs1999,Basar1999,Altman.Pourtallier.ea2000,Tijs2003}.
For example, a Nash equilibrium solution to a multi-objective optimization
problem is said to be achieved if none of the players can benefit
from a unilateral deviation from the equilibrium \cite{Nash1951}.
Thus, Nash equilibrium solutions provide a secure set of strategies
in the sense that none of the players have an incentive to diverge
from their equilibrium policy. Hence, Nash equilibrium has been a
widely used solution concept in differential game-based control techniques.
Online real-time solutions to differential games with centralized
objectives are presented in results such as \cite{Vamvoudakis2010a,Vrabie2010,Johnson2011a,Vamvoudakis2011,Lin.Cassandras2015};
however, since these results solve problems with centralized objectives
(i.e., each agent minimizes or maximizes a cost function that penalizes
the states of all the agents in the network), they are not applicable
for a network of agents with independent decentralized objectives
(i.e., each agent minimizes or maximizes a cost function that penalizes
only the error states corresponding to itself).

In this paper, the objective is to obtain an online forward-in-time
feedback-Nash equilibrium solution (cf. \cite{Case1969,Starr.Ho1969,Starr1969,Friedman1971,Bressan.Priuli2006,Bressan2011})
to an infinite-horizon formation tracking problem, where each agent
desires to follow a mobile leader while the group maintains a desired
formation. The agents try to minimize cost functions that penalize
their own formation tracking errors and their own control efforts.

Various methods have been developed to solve optimal tracking problems
for linear systems. %
In \cite{Vamvoudakis.Lewis.ea2012a,Wang.Xin2013,Zhang.Feng.ea2015,Ghosh.Lee2015},
optimal controllers are developed to cooperatively control agents
with linear dynamics. In \cite{Lin2014}, a differential game-based
approach is developed for unmanned aerial vehicles to achieve distributed
Nash strategies. In \cite{Semsar-Kazerooni.Khorasani2008}, an optimal
consensus algorithm is developed for a cooperative team of agents
with linear dynamics using only partial information. 

For nonlinear systems, a MPC-based approach is presented in \cite{Shim.Kim.ea2003};
however, no stability or convergence analysis is presented. A stable
distributed MPC-based approach is presented in \cite{Magni.Scattolini2006}
for nonlinear discrete-time systems with known nominal dynamics. Asymptotic
stability is proved without any interaction between the nodes; however,
a nonlinear optimal control problem needs to be solved at every iteration
to implement the controller. An optimal tracking approach for formation
control is presented in \cite{Heydari.Balakrishnan2012} using single
network adaptive critics where the value function is learned offline.
Recently, a leader-based consensus algorithm is developed in \cite{Zhang.Zhang.ea2015}
where exact model of the system dynamics is utilized, and convergence
to optimality is obtained under a persistence of excitation condition.

For multi-agent problems with decentralized objectives, the desired
action by an individual agent depends on the actions and the resulting
trajectories of its neighbors; hence, the error system for each agent
is a complex nonautonomous dynamical system. Nonautonomous systems,
in general, have non-stationary value functions. Since non-stationary
functions are difficult to approximate using parameterized function
approximation schemes such as neural networks (NNs), designing optimal
policies for nonautonomous systems is challenging.

Since the external influence from neighbors renders the dynamics of
each agent nonautonomous, optimization in a network of agents presents
challenges similar to optimal tracking problems. Using insights gained
from the authors' previous work on optimal tracking problems \cite{Kamalapurkar.Andrews.eatoappear},
this paper develops a model-based RL technique to generate feedback-Nash
equilibrium policies online, for agents in a network with cooperative
or competitive objectives. In particular, the network of agents is
separated into autonomous subgraphs, and the differential game is
solved separately on each subgraph. 

The primary contribution of this paper is the formulation and online
approximate feedback-Nash equilibrium solution of an optimal network
formation tracking problem. A relative control error minimization
technique is introduced to facilitate the formulation of a feasible
infinite-horizon total-cost differential graphical game. Dynamic programming-based
feedback-Nash equilibrium solution of the differential graphical game
is facilitated via the development of a set of coupled Hamilton-Jacobi
(HJ) equations. The developed approximate feedback-Nash equilibrium
solution is analyzed using a Lyapunov-based stability analysis to
demonstrate ultimately bounded formation tracking in the presence
of uncertainties.

\section{Notation}

Throughout the paper, $\mathbb{R}^{n}$ denotes $n-$dimensional Euclidean
space, $\mathbb{R}_{>a}$ denotes the set of real numbers strictly
greater than $a\in\mathbb{R}$, and $\mathbb{R}_{\geq a}$ denotes
the set of real numbers greater than or equal to $a\in\mathbb{R}$.
Unless otherwise specified, the domain of all the functions is assumed
to be $\mathbb{R}_{\geq0}$. Functions with domain $\mathbb{R}_{\geq0}$
are defined by abuse of notation using only their image. For example,
the function $x:\mathbb{R}_{\geq0}\to\mathbb{R}^{n}$ is defined by
abuse of notation as $x\in\mathbb{R}^{n}$. By abuse of notation,
the state variables are also used to denote state trajectories. For
example, the state variable $x$ in the equation $\dot{x}=f\left(x\right)+u$
is also used as $x\left(t\right)$ to denote the state trajectory,
i.e., the general solution $x:\mathbb{R}_{\geq0}\to\mathbb{R}^{n}$
to $\dot{x}=f\left(x\right)+u$ evaluated at time $t$. Unless otherwise
specified, all the mathematical quantities are assumed to be time-varying.
Unless otherwise specified, an equation of the form $g\left(x\right)=f+h\left(y,t\right)$
is interpreted as $g\left(x\left(t\right)\right)=f\left(t\right)+h\left(y\left(t\right),t\right)$
for all $t\in\mathbb{R}_{\geq0}$, and a definition of the form $g\left(x,y\right)\triangleq f\left(y\right)+h\left(x\right)$
for functions $g:A\times B\to C$, $f:B\to C$ and $h:A\to C$ is
interpreted as $g\left(x,y\right)\triangleq f\left(y\right)+h\left(x\right),\:\forall\left(x,y\right)\in A\times B$.
The total derivative $\frac{\partial f\left(x\right)}{\partial x}$
is denoted by $\nabla f$ and the partial derivative $\frac{\partial f\left(x,y\right)}{\partial x}$
is denoted by $\nabla_{x}f\left(x,y\right)$. An $n\times n$ identity
matrix is denoted by $I_{n}$, $n\times m$ matrices of zeros and
ones are denoted by $\mathbf{0}_{n\times m}$ and $\mathbf{1}_{n\times m}$,
respectively, and $\mathbf{1}_{S}$ denotes the indicator function
of the set $S$.

\section{Graph Theory Preliminaries}

Consider a set of $N$ autonomous agents moving in the state space
$\mathbb{R}^{n}$. The control objective is for the agents to maintain
a desired formation with respect to a leader. The state of the leader
is denoted by $x_{0}\in\mathbb{R}^{n}$. The agents are assumed to
be on a network with a fixed communication topology modeled as a static
directed graph (i.e. digraph). 

Each agent forms a node in the digraph. The set of all nodes excluding
the leader is denoted by $\mathcal{N}=\left\{ 1,\cdots N\right\} $
and the leader is denoted by node 0. If node $i$ can receive information
from node $j$ then there exists a directed edge from the $j\textsuperscript{th}$
to the $i$\textsuperscript{th} node of the digraph, denoted by the
ordered pair $\left(j,i\right)$. Let $E$ denote the set of all edges.
Let there be a positive weight $a_{ij}\in\mathbb{R}$ associated with
each edge $\left(j,i\right)$. Note that $a_{ij}\neq0$ if and only
if $\left(j,i\right)\in E.$ The digraph is assumed to have no repeated
edges, i.e., $\left(i,i\right)\notin E,\forall i$, which implies
$a_{ii}=0,\forall i$. The neighborhood sets of node $i$ are denoted
by $\mathcal{N}_{-i}$ and $\mathcal{N}_{i}$, defined as $\mathcal{N}_{-i}\triangleq\left\{ j\in\mathcal{N}\mid\left(j,i\right)\in E\right\} $
and $\mathcal{N}_{i}\triangleq\mathcal{N}_{-i}\cup\left\{ i\right\} $.

To streamline the analysis, an adjacency matrix $\mathcal{A}\in\mathbb{R}^{N\times N}$
is defined as $\mathcal{A}\triangleq\left[a_{ij}\mid i,j\in\mathcal{N}\right]$,
a diagonal pinning gain matrix $\mathcal{A}_{0}\in\mathbb{R}^{N\times N}$
is defined as $\mathcal{A}_{0}\triangleq\mbox{diag}\left(\left[a_{10},\cdots,a_{N0}\right]\right)$,
an in-degree matrix $\mathcal{D}\in\mathbb{R}^{N\times N}$ is defined
as $\mathcal{D}\triangleq\mbox{diag}\left(d_{i}\right),$ where $d_{i}\triangleq\sum_{j\in\mathcal{N}_{i}}a_{ij}$,
and a graph Laplacian matrix $\mathcal{L}\in\mathbb{R}^{N\times N}$
is defined as $\mathcal{L}\triangleq\mathcal{D}-\mathcal{A}$. The
graph is assumed to have a spanning tree, i.e., given any node $i$,
there exists a directed path from the leader $0$ to node $i$. A
node $j$ is said to be an extended neighbor of node $i$ if there
exists a directed path from node $j$ to node $i$. The extended neighborhood
set of node $i$, denoted by $\mathcal{S}_{-i}$, is defined as the
set of all extended neighbors of node $i.$ Formally, $\mathcal{S}_{-i}\triangleq\{j\in\mathcal{N}\mid j\neq i\wedge\exists\kappa\leq N,\:\left\{ j_{1},\cdots j_{\kappa}\right\} \subset\mathcal{N}\mid\left\{ \left(j,j_{1}\right),\left(j_{1},j_{2}\right),\cdots,\left(j_{\kappa},i\right)\right\} \subset2^{E}\}$.
Let $\mathcal{S}_{i}\triangleq\mathcal{S}_{-i}\cup\left\{ i\right\} $,
and let the edge weights be normalized such that $\sum_{j}a_{ij}=1$
for all $i\in\mathcal{N}$. Note that the sub-graphs are nested in
the sense that $\mathcal{S}_{j}\subseteq\mathcal{S}_{i}$ for all
$j\in\mathcal{S}_{i}$.

\section{Problem Formulation}

The state $x_{i}\in\mathbb{R}^{n}$ of each agent evolves according
to the control affine dynamics
\begin{equation}
\dot{x}_{i}=f_{i}\left(x_{i}\right)+g_{i}\left(x_{i}\right)u_{i},\label{eq:CLNNDyn}
\end{equation}
where $u_{i}\in\mathbb{R}^{m_{i}}$ denotes the control input, and
$f_{i}:\mathbb{R}^{n}\to\mathbb{R}^{n}$ and $g_{i}:\mathbb{R}^{n}\to\mathbb{R}^{n\times m_{i}}$
are locally Lipschitz continuous functions. 
\begin{assumption}
\label{ass:CLNNLeader}The dynamics of the leader are described by
$\dot{x}_{0}=f_{0}\left(x_{0}\right),$ where $f_{0}:\mathbb{R}^{n}\to\mathbb{R}^{n}$
is a locally Lipschitz continuous function. The function $f_{0}$,
and the initial condition $x_{0}\left(t_{0}\right)$ are selected
such that the trajectory $x_{0}\left(t\right)$ is uniformly bounded
for all $t\in\mathbb{R}_{\geq t_{0}}$.

The control objective is for the agents to maintain a predetermined
formation (with respect to an inertial reference frame) around the
leader while minimizing their own cost functions. For all $i\in\mathcal{N}$,
the $i\textsuperscript{th}$ agent is aware of its constant desired
relative position $x_{dij}\in\mathbb{R}^{n}$ with respect to all
its neighbors $j\in\mathcal{N}_{-i}$, such that the desired formation
is realized when $x_{i}-x_{j}\to x_{dij}$ for all $i,j\in\mathcal{N}$.\footnote{The vectors $x_{dij}$ are assumed to be fixed in an inertial reference
frame, i.e., the final desired formation is rigid and its motion in
an inertial reference frame can be described as pure translation.} To facilitate the control design, the formation is expressed in terms
of a set of constant vectors $\left\{ x_{di0}\in\mathbb{R}^{n}\right\} _{i\in\mathcal{N}}$
where each $x_{di0}$ denotes the constant final desired position
of agent $i$ with respect to the leader. The vectors $\left\{ x_{di0}\right\} _{i\in\mathcal{N}}$
are unknown to the agents not connected to the leader, and the known
desired inter agent relative position can be expressed in terms of
$\left\{ x_{di0}\right\} _{i\in\mathcal{N}}$ as $x_{dij}=x_{di0}-x_{dj0}$.
The control objective is thus satisfied when $x_{i}\to x_{di0}+x_{0}$
for all $i\in\mathcal{N}$. To quantify the objective, local neighborhood
tracking error signals are defined as 
\begin{equation}
e_{i}=\sum_{j\in\left\{ 0\right\} \cup\mathcal{N}_{-i}}a_{ij}\left(\left(x_{i}-x_{j}\right)-x_{dij}\right).\label{eq:CLNNe}
\end{equation}

To facilitate the analysis, the error signals in (\ref{eq:CLNNe})
are expressed in terms of the unknown leader-relative desired positions
as 
\begin{equation}
e_{i}=\sum_{j\in\left\{ 0\right\} \cup\mathcal{N}_{-i}}a_{ij}\left(\left(x_{i}-x_{di0}\right)-\left(x_{j}-x_{dj0}\right)\right).\label{eq:CLNNeUnk}
\end{equation}
Stacking the error signals in a vector $\mathcal{E}\triangleq\left[\begin{array}[t]{cccc}
e_{1}^{T}, & e_{2}^{T}, & \cdots, & e_{N}^{T}\end{array}\right]^{T}\in\mathbb{R}^{nN}$ the equation in (\ref{eq:CLNNeUnk}) can be expressed in a matrix
form 
\begin{equation}
\mathcal{E}=\left(\left(\mathcal{L}+\mathcal{A}_{0}\right)\otimes I_{n}\right)\left(\mathcal{X}-\mathcal{X}_{d}-\mathcal{X}_{0}\right),\label{eq:CLNNBigE}
\end{equation}
where $\mathcal{X}=$ $\left[x_{1}^{T},\right.$ $x_{2}^{T},$ $\cdots,$
$\left.x_{N}^{T}\right]^{T}$$\in\mathbb{R}^{nN}$, $\mathcal{X}_{d}=$
$\left[x_{d10}^{T},\right.$ $x_{d20}^{T},$ $\cdots,$ $\left.x_{dN0}^{T}\right]^{T}$$\in\mathbb{R}^{nN}$,
$\mathcal{X}_{0}=$ $\left[x_{0}^{T},\right.$ $x_{0}^{T},$ $\cdots,$
$\left.x_{0}^{T}\right]^{T}$$\in\mathbb{R}^{nN}$, and $\otimes$
denotes the Kronecker product. Using (\ref{eq:CLNNBigE}), it can
be concluded that provided the matrix $\left(\left(\mathcal{L}+\mathcal{A}_{0}\right)\otimes I_{n}\right)\in\mathbb{R}^{nN\times nN}$
is nonsingular, $\left\Vert \mathcal{E}\right\Vert \to0$ implies
$x_{i}\to x_{di0}+x_{0}$ for all $i\in\mathcal{N}$, and hence, the
satisfaction of control objective. The matrix $\left(\left(\mathcal{L}+\mathcal{A}_{0}\right)\otimes I_{n}\right)$
is nonsingular provided the graph has a spanning tree with the leader
at the root \cite{Khoo.Xie2009}. To facilitate the formulation of
an optimization problem, the following section explores the functional
dependence of the state-value functions for the network of agents. 
\end{assumption}

\subsection{\label{subsec:CLNNElements-of-Value-Function}Elements of the value
function}

The dynamics for the open-loop neighborhood tracking error are %
$\dot{e}_{i}=\sum_{j\in\left\{ 0\right\} \cup\mathcal{N}_{-i}}a_{ij}\Bigl(f_{i}\left(x_{i}\right)+g_{i}\left(x_{i}\right)u_{i}-f_{j}\left(x_{j}\right)-g_{j}\left(x_{j}\right)u_{j}\Bigr).$
Under the temporary assumption that each controller $u_{i}$ is an
error-feedback controller, i.e. $u_{i}\left(t\right)=\hat{u}_{i}\left(e_{i}\left(t\right),t\right)$,
the error dynamics are expressed as $\dot{e}_{i}=\sum_{j\in\left\{ 0\right\} \cup\mathcal{N}_{-i}}a_{ij}\Bigl(f_{i}\left(x_{i}\right)+g_{i}\left(x_{i}\right)\hat{u}_{i}\left(e_{i},t\right)-f_{j}\left(x_{j}\right)-g_{j}\left(x_{j}\right)\hat{u}_{j}\left(e_{j},t\right)\Bigr).$
Thus, the error trajectory $\left\{ e_{i}\left(t\right)\right\} _{t=t_{0}}^{\infty}$
, where $t_{0}$ denotes the initial time, depends on $\hat{u}_{j}\left(e_{j}\left(t\right),t\right)$,
$\forall j\in\mathcal{N}_{i}$. Similarly, the error trajectory $\left\{ e_{j}\left(t\right)\right\} _{t=t_{0}}^{\infty}$
depends on $\hat{u}_{k}\left(e_{k}\left(t\right),t\right),\forall k\in\mathcal{N}_{j}$.
Recursively, the trajectory $\left\{ e_{i}\left(t\right)\right\} _{t=t_{0}}^{\infty}$
depends on $\hat{u}_{j}\left(e_{j}\left(t\right),t\right)$, and hence,
on $e_{j}\left(t\right),\forall j\in\mathcal{S}_{i}$. Thus, even
if the controller for each agent is restricted to use local error
feedback, the resulting error trajectories are interdependent. In
particular, a change in the initial condition of one agent in the
extended neighborhood causes a change in the error trajectories corresponding
to all the extended neighbors. Consequently, the value function corresponding
to an infinite-horizon optimal control problem where each agent tries
to minimize $\intop_{t_{0}}^{\infty}\left(Q\left(e_{i}\left(\tau\right)\right)+R\left(u_{i}\left(\tau\right)\right)\right)d\tau$,
where $Q:\mathbb{R}^{n}\to\mathbb{R}$ and $R:\mathbb{R}^{m_{i}}\to\mathbb{R}$
are positive definite functions, is dependent on the error states
of all the extended neighbors.%

Since the steady-state controllers required for formation tracking
are generally nonzero, quadratic total-cost optimal control problems
result in infinite costs, and hence, are infeasible. In the following
section, relative steady-state controllers are derived to facilitate
the formulation of a feasible optimal control problem.

\subsection{Optimal formation tracking problem}

When the agents are perfectly tracking the desired trajectory in the
desired formation, even though the states of all the agents are different,
the time-derivatives of the states of all the agents are identical.
Hence, in steady state, the control signal applied by each agent must
be such that the time derivatives of the states corresponding to the
set of extended neighbors are identical. In particular, the relative
control signal $u_{ij}\in\mathbb{R}^{m_{i}}$ that will keep node
$i$ in its desired relative position with respect to node $j\in\mathcal{S}_{-i}$,
i.e., $x_{i}=x_{j}+x_{dij}$, must be such that the time derivative
of $x_{i}$ is the same as the time derivative of $x_{j}$. Using
the dynamics of the agents from (\ref{eq:CLNNDyn}), and substituting
the desired relative positions $x_{j}+x_{dij}$ for the states $x_{i}$,
the relative control signals $u_{ij}$ must satisfy 
\begin{equation}
f_{i}\left(x_{j}+x_{dij}\right)+g_{i}\left(x_{j}+x_{dij}\right)u_{ij}=\dot{x}_{j}.\label{eq:CLNNRelDyn}
\end{equation}
The relative steady-state control signals can be expressed in an explicit
form provided the following assumption is satisfied. 
\begin{assumption}
\label{ass:Pseudo} The matrix $g_{i}\left(x\right)$ is full rank
for all $i\in\mathcal{N}$ and for all $x\in\mathbb{R}^{n}$; furthermore,
the relative steady-state control signal expressed as%
{} $u_{ij}=f_{ij}\left(x_{j}\right)+g_{ij}\left(x_{j}\right)u_{j},$
satisfies (\ref{eq:CLNNRelDyn}) along the desired trajectory, where
$f_{ij}\left(x_{j}\right)\triangleq g_{i}^{+}\left(x_{j}+x_{dij}\right)\left(f_{j}\left(x_{j}\right)-f_{i}\left(x_{j}+x_{dij}\right)\right)\in\mathbb{R}^{m_{i}}$,
$g_{ij}\left(x_{j}\right)\triangleq g_{i}^{+}\left(x_{j}+x_{dij}\right)g_{j}\left(x_{j}\right)\in\mathbb{R}^{m_{i}\times m_{j}}$,
$g_{0}\left(x\right)\triangleq0$ for all $x\in\mathbb{R}^{n}$, $u_{i0}\equiv0$
for all $i\in\mathcal{N}$, and $g_{i}^{+}\left(x\right)$ denotes
a pseudoinverse of the matrix $g_{i}\left(x\right)$ for all $x\in\mathbb{R}^{n}$
and for all $i\in\mathcal{N}$.
\end{assumption}
Assumption \ref{ass:Pseudo} places restrictions on the control-effectiveness
matrices. The matrices $g_{i}\left(x\right)$ are full rank for a
large class of systems including, but not limited to, kinematic wheels
and fully actuated Euler-Lagrange systems with invertible inertia
matrices. The second part of Assumption 2 requires the existence of
a feedback controller that can keep the system on the desired trajectory
if the system starts on the desired trajectory. This assumption depends
on the systems, the network, the desired formation, and the desired
trajectory; hence, insights into its satisfaction are hard to obtain
in general. The satisfaction of this assumption needs to be verified
on a case-by-case basis. For example, consider a kinematic wheel modeled
as 
\begin{equation}
\dot{x}=g\left(x\right)u,\quad g\left(x\right)=\begin{bmatrix}\cos\left(x_{3}\right) & 0\\
\sin\left(x_{3}\right) & 0\\
0 & 1
\end{bmatrix}.\label{eq:CLNNWheel}
\end{equation}
In this case, provided the formation satisfies $x_{dij}\left(3\right)=0$,
that is, the target formation is such that all the kinematic wheels
have the same steering angle, the functions $f_{ij}$ and $g_{ij}$
can be computed as $f_{ij}=0,$ and $g_{ij}=I_{2}$. The relative
steady-state control is then $u_{ij}=u_{j}$, which satisfies $g\left(x_{j}+x_{dij}\right)u_{j}=\dot{x}_{j}$,
and hence, Assumption \ref{ass:Pseudo} holds.

To facilitate the formulation of an optimal formation tracking problem,
define the control errors $\mu_{i}\in\mathbb{R}^{m_{i}}$ as 
\begin{equation}
\mu_{i}\triangleq\sum_{j\in\mathcal{N}_{-i}\cup\left\{ 0\right\} }a_{ij}\left(u_{i}-u_{ij}\right).\label{eq:CLNNmu_i}
\end{equation}
The control errors $\left\{ \mu_{i}\right\} $ are treated as the
design variables in the remainder of this paper. Since the control
errors $\left\{ \mu_{i}\right\} $ are designed and the controllers
$\left\{ u_{i}\right\} $ are implemented in practice, it is essential
to invert the relationship in (\ref{eq:CLNNmu_i}). To facilitate
the inversion, let $\mathcal{S}_{i}^{o}\triangleq\left\{ 1,\cdots,s_{i}\right\} $,
where $s_{i}\triangleq\left|\mathcal{S}_{i}\right|$. Let $\lambda_{i}:\mathcal{S}_{i}^{o}\to\mathcal{S}_{i}$
be a bijective map such that $\lambda_{i}\left(1\right)=i$. For notational
brevity, let $\left(\cdot\right)_{\mathcal{S}_{i}}$ denote the concatenated
vector $\left[\left(\cdot\right)_{\lambda_{i}^{1}}^{T},\left(\cdot\right)_{\lambda_{i}^{2}}^{T},\cdots,\left(\cdot\right)_{\lambda_{i}^{s_{i}}}^{T}\right]^{T}$,
let $\left(\cdot\right)_{\mathcal{S}_{-i}}$ denote the concatenated
vector $\left[\left(\cdot\right)_{\lambda_{i}^{2}}^{T},\cdots,\left(\cdot\right)_{\lambda_{i}^{s_{i}}}^{T}\right]^{T}$,
let $\sum^{i}$ denote $\sum_{j\in\mathcal{N}_{-i}\cup\left\{ 0\right\} }$,
let $\lambda_{i}^{j}$ denote $\lambda_{i}\left(j\right)$, let $\mathcal{E}_{i}\triangleq\left[e_{\mathcal{S}_{i}}^{T},x_{\lambda_{i}^{1}}^{T}\right]^{T}\in\mathbb{R}^{n\left(s_{i}+1\right)}$,
and let $\mathcal{E}_{-i}\triangleq\left[e_{\mathcal{S}_{-i}}^{T},x_{\lambda_{i}^{1}}^{T}\right]^{T}\in\mathbb{R}^{ns_{i}}$.
Then, the control error vectors $\mu_{\mathcal{S}_{i}}\in\mathbb{R}^{\sum_{k\in\mathcal{S}_{i}}m_{k}}$
can be expressed as%
\begin{equation}
\mu_{\mathcal{S}_{i}}=\mathscr{L}_{gi}\left(\mathcal{E}_{i}\right)u_{\mathcal{S}_{i}}-F_{i}\left(\mathcal{E}_{i}\right),\label{eq:CLNNmu_S_i}
\end{equation}
where the matrices $\mathscr{L}_{gi}:\mathbb{R}^{n\left(s_{i}+1\right)}\to\mathbb{R}^{\sum_{k\in\mathcal{S}_{i}}m_{k}\times\sum_{k\in\mathcal{S}_{i}}m_{k}}$
are defined by 
\[
\left[\mathscr{L}_{gi}\left(\mathcal{E}_{i}\right)\right]_{kl}=\begin{cases}
-a_{\lambda_{i}^{k}\lambda_{i}^{l}}g_{\lambda_{i}^{k}\lambda_{i}^{l}}\left(x_{\lambda_{i}^{l}}\right), & \forall l\neq k,\\
\sideset{}{^{\lambda_{i}^{k}}}\sum a_{\lambda_{i}^{k}j}I_{m_{\lambda_{i}^{k}}}, & \forall l=k,
\end{cases}
\]
where $k,l=1,2,\cdots,s_{i}$,%
{} and $F_{i}:\mathbb{R}^{n\left(s_{i}+1\right)}\to\mathbb{R}^{\sum_{k\in\mathcal{S}_{i}}m_{k}}$
are defined as 
\[
F_{i}\!\left(\!\mathcal{E}_{i}\!\right)\!\triangleq\!\left[\!\sideset{}{^{i}}\sum\!\!a_{\lambda_{i}^{1}j}f_{\lambda_{i}^{1}j}^{T}\!\left(\!x_{j}\!\right)\!,\!\cdots\!,\!\!\sideset{}{^{\lambda_{i}^{s_{i}}}}\sum\!\!a_{\lambda_{i}^{s_{i}}j}f_{\lambda_{i}^{s_{i}}j}^{T}\!\left(\!x_{j}\!\right)\!\right]^{T}.
\]

\begin{assumption}
\label{ass:CLNNLgiInvertible}The matrix $\mathscr{L}_{gi}\left(\mathcal{E}_{i}\left(t\right)\right)$
is invertible for all $t\in\mathbb{R}$ and for all $i\in\mathcal{N}$.
\end{assumption}
Assumption \ref{ass:CLNNLgiInvertible} is a controllability-like
condition. Intuitively, Assumption \ref{ass:CLNNLgiInvertible} requires
the control effectiveness matrices to be compatible to ensure the
existence of relative control inputs that allow the agents to follow
the desired trajectory in the desired formation. Assumption \ref{ass:CLNNLgiInvertible}
depends on the systems, the network, the desired formation, and the
desired trajectory; hence, insights into its satisfaction are hard
to obtain in general. The satisfaction of this assumption needs to
be verified on a case-by-case basis. For example, consider the kinematic
wheel in (\ref{eq:CLNNWheel}). Provided the formation satisfies $x_{dij}\left(3\right)=0$,
that is, the target formation is such that all the kinematic wheels
have the same steering angle, we have $g_{ij}=I_{2},$ and hence,
the matrices $\mathscr{L}_{gi}$ are given by 
\[
\left[\mathscr{L}_{gi}\left(\mathcal{E}_{i}\right)\right]_{kl}=\begin{cases}
-a_{\lambda_{i}^{k}\lambda_{i}^{l}}I_{2}, & \forall l\neq k,\\
\sideset{}{^{\lambda_{i}^{k}}}\sum a_{\lambda_{i}^{k}j}I_{2}, & \forall l=k,
\end{cases}
\]
It can be shown that $\mathscr{L}_{gi}=\mathcal{L}_{\mathcal{S}_{i}}\otimes I_{2}$,
where $\mathcal{L}_{\mathcal{S}_{i}}$ denotes the Laplacian matrix
corresponding to the subgraph $\mathcal{S}_{i}$. Hence, the graph
connectivity condition ensures that the matrices $\mathscr{L}_{gi}$
are invertible, and in this specific case, Assumption \ref{ass:CLNNLgiInvertible}
holds. 

Using Assumption \ref{ass:CLNNLgiInvertible}, the control vectors
can be expressed as 
\begin{equation}
u_{\mathcal{S}_{i}}=\mathscr{L}_{gi}^{-1}\left(\mathcal{E}_{i}\right)\mu_{\mathcal{S}_{i}}+\mathscr{L}_{gi}^{-1}\left(\mathcal{E}_{i}\right)F_{i}\left(\mathcal{E}_{i}\right).\label{eq:CLNNU_S_i}
\end{equation}
Let $\mathscr{L}_{gi}^{k}$ denote the $\left(\lambda_{i}^{-1}\left(k\right)\right)$\textsuperscript{th}
block row of $\mathscr{L}_{gi}^{-1}$. Then, the controllers $u_{i}$
can be implemented as 
\begin{equation}
u_{i}=\mathscr{L}_{gi}^{i}\left(\mathcal{E}_{i}\right)\mu_{\mathcal{S}_{i}}+\mathscr{L}_{gi}^{i}\left(\mathcal{E}_{i}\right)F_{i}\left(\mathcal{E}_{i}\right),\label{eq:CLNNu_i}
\end{equation}
and for any $j\in\mathcal{N}_{-i}$, 
\begin{equation}
u_{j}=\mathscr{L}_{gi}^{j}\left(\mathcal{E}_{i}\right)\mu_{\mathcal{S}_{i}}+\mathscr{L}_{gi}^{j}\left(\mathcal{E}_{i}\right)F_{i}\left(\mathcal{E}_{i}\right).\label{eq:CLNNu_j}
\end{equation}
Using (\ref{eq:CLNNu_i}) and (\ref{eq:CLNNu_j}), the error and the
state dynamics for the agents can be represented as %
\begin{equation}
\dot{e}_{i}=\mathscr{F}_{i}\left(\mathcal{E}_{i}\right)+\mathscr{G}_{i}\left(\mathcal{E}_{i}\right)\mu_{\mathcal{S}_{i}},\label{eq:CLNNe_iDot}
\end{equation}
and %
\begin{equation}
\dot{x}_{i}=\mathcal{F}_{i}\left(\mathcal{E}_{i}\right)+\mathcal{G}_{i}\left(\mathcal{E}_{i}\right)\mu_{\mathcal{S}_{i}},\label{eq:CLNNx_iDot}
\end{equation}
where 
\begin{multline*}
\mathscr{F}_{i}\left(\mathcal{E}_{i}\right)\triangleq\!\!\sideset{}{^{i}}\sum a_{ij}g_{i}\left(x_{i}\right)\mathscr{L}_{gi}^{i}\left(\mathcal{E}_{i}\right)F_{i}\left(\mathcal{E}_{i}\right)\!-\!\!\sideset{}{^{i}}\sum a_{ij}f_{j}\left(x_{j}\right)\\
-\!\!\sideset{}{^{i}}\sum a_{ij}g_{j}\left(x_{j}\right)\mathscr{L}_{gi}^{j}\left(\mathcal{E}_{i}\right)F_{i}\left(\mathcal{E}_{i}\right)+\!\!\sideset{}{^{i}}\sum a_{ij}f_{i}\left(x_{i}\right),
\end{multline*}
\[
\mathscr{G}_{i}\left(\mathcal{E}_{i}\right)\triangleq\!\!\sideset{}{^{i}}\sum a_{ij}\left(g_{i}\left(x_{i}\right)\mathscr{L}_{gi}^{i}\left(\mathcal{E}_{i}\right)-g_{j}\left(x_{j}\right)\mathscr{L}_{gi}^{j}\left(\mathcal{E}_{i}\right)\right),
\]
\[
\mathcal{F}_{i}\left(\mathcal{E}_{i}\right)\triangleq f_{i}\left(x_{i}\right)+g_{i}\left(x_{i}\right)\mathscr{L}_{gi}^{i}\left(\mathcal{E}_{i}\right)F_{i}\left(\mathcal{E}_{i}\right),
\]
 and $\mathcal{G}_{i}\left(\mathcal{E}_{i}\right)\triangleq g_{i}\left(x_{i}\right)\mathscr{L}_{gi}^{i}\left(\mathcal{E}_{i}\right)$.

Let $h_{ei}^{\overline{\mu}_{i},\overline{\mu}_{\mathcal{S}_{-i}}}\left(t,t_{0},\mathcal{E}_{i0}\right)$
and $h_{xi}^{\overline{\mu}_{i},\overline{\mu}_{\mathcal{S}_{-i}}}\left(t,t_{0},\mathcal{E}_{i0}\right)$
denote the trajectories of (\ref{eq:CLNNe_iDot}) and (\ref{eq:CLNNx_iDot}),
respectively, with the initial time $t_{0}$, initial condition $\mathcal{E}_{i}\left(t_{0}\right)=\mathcal{E}_{i0}$,
and policies $\overline{\mu}_{j}:\mathbb{R}^{n\left(s_{i}+1\right)}\to\mathbb{R}^{m_{i}},\:j\in\mathcal{S}_{i}$,
and let $\mathcal{H}_{i}\triangleq\left[\left(h_{e}\right)_{\mathcal{S}_{i}}^{T},h_{x\lambda_{i}^{1}}^{T}\right]^{T}$.
Define the cost functionals 
\begin{equation}
J_{i}\left(e_{i}\left(\cdot\right),\mu_{i}\left(\cdot\right)\right)\triangleq\intop_{0}^{\infty}r_{i}\left(e_{i}\left(\sigma\right),\mu_{i}\left(\sigma\right)\right)\textnormal{d}\sigma\label{eq:CLNNJi}
\end{equation}
where $r_{i}:\mathbb{R}^{n}\times\mathbb{R}^{m_{i}}\to\mathbb{R}_{\geq0}$
denote the local costs defined as $r_{i}\left(e_{i},\mu_{i}\right)\triangleq Q_{i}\left(e_{i}\right)+\mu_{i}^{T}R_{i}\mu_{i},$
where $Q_{i}:\mathbb{R}^{n}\to\mathbb{R}_{\geq0}$ are positive definite
functions, and $R_{i}\in\mathbb{R}^{m_{i}\times m_{i}}$ are constant
positive definite matrices. The objective of each agent is to minimize
the cost functional in (\ref{eq:CLNNJi}). To facilitate the definition
of a feedback-Nash equilibrium solution, define the value functions
$V_{i}:\mathbb{R}^{n\left(s_{i}+1\right)}\to\mathbb{R}_{\geq0}$ as
\begin{multline}
V_{i}^{\overline{\mu}_{i},\overline{\mu}_{\mathcal{S}_{-i}}}\left(\mathcal{E}_{i}\right)\triangleq\\
\intop_{t}^{\infty}r_{i}\left(h_{ei}^{\overline{\mu}_{i},\overline{\mu}_{\mathcal{S}_{-i}}}\left(\sigma,t,\mathcal{E}_{i}\right),\overline{\mu}_{i}\left(\mathcal{H}_{i}^{\overline{\mu}_{i},\overline{\mu}_{\mathcal{S}_{-i}}}\left(\sigma,t,\mathcal{E}_{i}\right)\right)\right)\textnormal{d}\sigma,\label{eq:CLNNV_i}
\end{multline}
where $V_{i}^{\overline{\mu}_{i},\overline{\mu}_{\mathcal{S}_{-i}}}\left(\mathcal{E}_{i}\right)$
denotes the total cost-to-go for Agent $i$ under the policies $\overline{\mu}_{\mathcal{S}_{i}}$,
when the sub-graph $\mathcal{S}_{i}$ starts from the state $\mathcal{E}_{i}$.
Note that the value functions in (\ref{eq:CLNNV_i}) are time-invariant
because the dynamical systems $\left\{ \dot{e}_{j}=\mathscr{F}_{j}\left(\mathcal{E}_{i}\right)+\mathscr{G}_{j}\left(\mathcal{E}_{i}\right)\mu_{\mathcal{S}_{j}}\right\} _{j\in\mathcal{S}_{i}}$
and $\dot{x}_{i}=\mathcal{F}_{i}\left(\mathcal{E}_{i}\right)+\mathcal{G}_{i}\left(\mathcal{E}_{i}\right)\mu_{\mathcal{S}_{i}}$
together form an autonomous dynamical system. 

A graphical feedback-Nash equilibrium solution within the subgraph
$\mathcal{S}_{i}$ is defined as the tuple of policies $\left\{ \mu_{j}^{*}:\mathbb{R}^{n\left(s_{j}+1\right)}\to\mathbb{R}^{m_{j}}\right\} _{j\in\mathcal{S}_{i}}$
such that the value functions in (\ref{eq:CLNNV_i}) satisfy 
\[
V_{j}^{*}\left(\mathcal{E}_{j}\right)\triangleq V_{j}^{\mu_{j}^{*},\mu_{\mathcal{S}_{-j}}^{*}}\left(\mathcal{E}_{j}\right)\leq V_{j}^{\overline{\mu}_{j},\mu_{\mathcal{S}_{-j}}^{*}}\left(\mathcal{E}_{j}\right),
\]
for all $j\in\mathcal{S}_{i}$, for all $\mathcal{E}_{i}\in\mathbb{R}^{n\left(s_{i}+1\right)}$
and for all admissible policies $\overline{\mu}_{j}$. Provided a
feedback-Nash equilibrium solution exists and the value functions
(\ref{eq:CLNNV_i}) are continuously differentiable for all $i\in\mathcal{N}$,
the feedback-Nash equilibrium value functions can be characterized
in terms of the following system of HJ equations:
\begin{multline}
\sum_{j\in\mathcal{S}_{i}}\nabla_{e_{j}}V_{i}^{*}\left(\mathcal{E}_{i}\right)\left(\mathscr{F}_{j}\left(\mathcal{E}_{i}\right)+\mathscr{G}_{j}\left(\mathcal{E}_{i}\right)\mu_{\mathcal{S}_{j}}^{*}\left(\mathcal{E}_{i}\right)\right)\\
+\nabla_{x_{i}}V_{i}^{*}\left(\mathcal{E}_{i}\right)\left(\mathcal{F}_{i}\left(\mathcal{E}_{i}\right)+\mathcal{G}_{i}\left(\mathcal{E}_{i}\right)\mu_{\mathcal{S}_{i}}^{*}\left(\mathcal{E}_{i}\right)\right)\\
+\overline{Q}_{i}\left(\mathcal{E}_{i}\right)+\mu_{i}^{*T}\left(\mathcal{E}_{i}\right)R_{i}\mu_{i}^{*}\left(\mathcal{E}_{i}\right)=0,\:\forall\mbox{\ensuremath{\mathcal{E}}}_{i}\in\mathbb{R}^{n\left(s_{i}+1\right)},\label{eq:CLNNHJB}
\end{multline}
where $\overline{Q}_{i}:\mathbb{R}^{n\left(s_{i}+1\right)}\to\mathbb{R}$
is defined as $\overline{Q}_{i}\left(\mathcal{E}_{i}\right)\triangleq Q_{i}\left(e_{i}\right)$.
\begin{thm}
\label{thm:CLNNNesSuf}Provided a feedback-Nash equilibrium solution
exists and that the value functions in (\ref{eq:CLNNV_i}) are continuously
differentiable, the system of HJ equations in (\ref{eq:CLNNHJB})
constitutes a necessary and sufficient condition for $\left\{ \mu_{j}^{*}:\mathbb{R}^{n\left(s_{j}+1\right)}\to\mathbb{R}^{m_{j}}\right\} _{j\in\mathcal{S}_{i}}$
to be a feedback-Nash equilibrium solution within the subgraph $\mathcal{S}_{i}$.
\end{thm}
\begin{IEEEproof}
Consider the cost functional in (\ref{eq:CLNNJi}), and assume that
all the extended neighbors of the $i\textsuperscript{th}$ agent follow
their feedback-Nash equilibrium policies. The value function corresponding
to any admissible policy $\overline{\mu}_{i}$ can be expressed as
\begin{multline*}
V_{i}^{\overline{\mu}_{i},\mu_{\mathcal{S}_{-i}}^{*}}\left(\left[e_{i}^{T},\:\mathcal{E}_{-i}^{T}\right]^{T}\right)=\\
\intop_{t}^{\infty}r_{i}\left(h_{ei}^{\overline{\mu}_{i},\mu_{\mathcal{S}_{-i}}^{*}}\left(\sigma,t,\mathcal{E}_{i}\right),\overline{\mu}_{i}\left(\mathcal{H}_{i}^{\overline{\mu}_{i},\mu_{\mathcal{S}_{-i}}^{*}}\left(\sigma,t,\mathcal{E}_{i}\right)\right)\right)\textnormal{d}\sigma.
\end{multline*}
Treating the dependence on $\mathcal{E}_{-i}$ as explicit time dependence
define 
\begin{equation}
\overline{V}_{i}^{\overline{\mu}_{i},\mu_{\mathcal{S}_{-i}}^{*}}\left(e_{i},t\right)\triangleq V_{i}^{\overline{\mu}_{i},\mu_{\mathcal{S}_{-i}}^{*}}\left(\left[e_{i}^{T},\:\mathcal{E}_{-i}^{T}\left(t\right)\right]^{T}\right),\label{eq:CLNNVbari}
\end{equation}
for all $e_{i}\in\mathbb{R}^{n}$ and for all $t\in\mathbb{R}_{\geq0}$.
Assuming that the optimal controller that minimizes (\ref{eq:CLNNJi})
when all the extended neighbors follow their feedback-Nash equilibrium
policies exists, and that the optimal value function $\overline{V}_{i}^{*}\triangleq\overline{V}_{i}^{\mu_{i}^{*},\mu_{\mathcal{S}_{-i}}^{*}}$
exists and is continuously differentiable, optimal control theory
for single objective optimization problems (cf. \cite{Liberzon2012})
can be used to derive the following necessary and sufficient condition
\begin{multline}
\frac{\partial\overline{V}_{i}^{*}\left(e_{i},t\right)}{\partial e_{i}}\left(\mathscr{F}_{i}\left(\mathcal{E}_{i}\right)+\mathscr{G}_{i}\left(\mathcal{E}_{i}\right)\mu_{\mathcal{S}_{i}}^{*}\left(\mathcal{E}_{i}\right)\right)+\frac{\partial\overline{V}_{i}^{*}\left(e_{i},t\right)}{\partial t}\\
+Q_{i}\left(e_{i}\right)+\mu_{i}^{*T}\left(\mathcal{E}_{i}\right)R_{i}\mu_{i}^{*}\left(\mathcal{E}_{i}\right)=0.\label{eq:CLNNHJB1}
\end{multline}
Using (\ref{eq:CLNNVbari}), the partial derivative with respect to
the state can be expressed as
\begin{equation}
\frac{\partial\overline{V}_{i}^{*}\left(e_{i},t\right)}{\partial e_{i}}=\frac{\partial V_{i}^{*}\left(\mathcal{E}_{i}\right)}{\partial e_{i}},\label{eq:CLNNPartialVi1}
\end{equation}
for all $e_{i}\in\mathbb{R}^{n}$ and for all $t\in\mathbb{R}_{\geq0}$,
and the partial derivative with respect to time can be expressed as
\begin{multline}
\frac{\partial\overline{V}_{i}^{*}\left(e_{i},t\right)}{\partial t}=\frac{\partial V_{i}^{*}\left(\mathcal{E}_{i}\right)}{\partial x_{i}}\left(\mathcal{F}_{i}\left(\mathcal{E}_{i}\right)+\mathcal{G}_{i}\left(\mathcal{E}_{i}\right)\mu_{\mathcal{S}_{i}}^{*}\left(\mathcal{E}_{i}\right)\right)\\
+\sum_{j\in\mathcal{S}_{-i}}\frac{\partial V_{i}^{*}\left(\mathcal{E}_{i}\right)}{\partial e_{j}}\left(\mathscr{F}_{j}\left(\mathcal{E}_{i}\right)+\mathscr{G}_{j}\left(\mathcal{E}_{i}\right)\mu_{\mathcal{S}_{j}}^{*}\left(\mathcal{E}_{i}\right)\right),\label{eq:CLNNPartialVi2}
\end{multline}
for all $e_{i}\in\mathbb{R}^{n}$ and for all $t\in\mathbb{R}_{\geq0}$.
Substituting (\ref{eq:CLNNPartialVi1}) and (\ref{eq:CLNNPartialVi2})
into (\ref{eq:CLNNHJB1}) and repeating the process for each $i$,
the system of HJ equations in (\ref{eq:CLNNHJB}) is obtained.
\end{IEEEproof}
Minimizing the HJ equations using the stationary condition, the feedback-Nash
equilibrium solution is expressed in the explicit form %
\begin{multline}
\mu_{i}^{*}\left(\mathcal{E}_{i}\right)=-\frac{1}{2}R_{i}^{-1}\sum_{j\in\mathcal{S}_{i}}\left(\mathscr{G}_{j}^{i}\left(\mathcal{E}_{i}\right)\right)^{T}\left(\nabla_{e_{j}}V_{i}^{*}\left(\mathcal{E}_{i}\right)\right)^{T}\\
-\frac{1}{2}R_{i}^{-1}\left(\mathcal{G}_{i}^{i}\left(\mathcal{E}_{i}\right)\right)^{T}\left(\nabla_{x_{i}}V_{i}^{*}\left(\mathcal{E}_{i}\right)\right)^{T},\label{eq:CLNNmu_i*}
\end{multline}
for all $\mathcal{E}_{i}\in\mathbb{R}^{n\left(s_{i}+1\right)}$, where
$\mathscr{G}_{j}^{i}\triangleq\mathscr{G}_{j}\frac{\partial\mu_{\mathcal{S}_{j}}^{*}}{\partial\mu_{i}^{*}}$,
and $\mathcal{G}_{i}^{i}\triangleq\mathcal{G}_{i}\frac{\partial\mu_{\mathcal{S}_{i}}^{*}}{\partial\mu_{i}^{*}}$.
Since an analytical solution of system of HJ equations in (\ref{eq:CLNNHJB})
is generally infeasible to obtain, the feedback-Nash value functions
and the feedback-Nash policies are approximated using parametric approximation
schemes $\hat{V}_{i}\left(\mathcal{E}_{i},\hat{W}_{ci}\right)$ and
$\hat{\mu}_{i}\left(\mathcal{E}_{i},\hat{W}_{ai}\right)$, respectively,
where $\hat{W}_{ci}\in\mathbb{R}^{L_{i}}$ and $\hat{W}_{ai}\in\mathbb{R}^{L_{i}}$
are parameter estimates. Substitution of the approximations $\hat{V}_{i}$
and $\hat{\mu}_{i}$ in (\ref{eq:CLNNHJB}) leads to a set of Bellman
errors (BEs) $\delta_{i}$ defined as
\begin{multline}
\delta_{i}\left(\mathcal{E}_{i},\hat{W}_{ci},\left(\hat{W}_{a}\right)_{\mathcal{S}_{i}}\right)\triangleq\hat{\mu}_{i}^{T}\left(\mathcal{E}_{i},\hat{W}_{ai}\right)R\hat{\mu}_{i}\left(\mathcal{E}_{i},\hat{W}_{ai}\right)\\
+\sum_{j\in\mathcal{S}_{i}}\nabla_{e_{j}}\hat{V}_{i}\left(\mathcal{E}_{i},\hat{W}_{ci}\right)\mathscr{G}_{j}\left(\mathcal{E}_{j}\right)\hat{\mu}_{\mathcal{S}_{j}}\left(\mathcal{E}_{j},\left(\hat{W}_{a}\right)_{\mathcal{S}_{j}}\right)\\
+\nabla_{x_{i}}\hat{V}_{i}\left(\mathcal{E}_{i},\hat{W}_{ci}\right)\left(\mathcal{F}_{i}\left(\mathcal{E}_{i}\right)+\mathcal{G}_{i}\left(\mathcal{E}_{i}\right)\hat{\mu}_{\mathcal{S}_{i}}\left(\mathcal{E}_{i},\left(\hat{W}_{a}\right)_{\mathcal{S}_{i}}\right)\right)\\
+\sum_{j\in\mathcal{S}_{i}}\nabla_{e_{j}}\hat{V}_{i}\left(\mathcal{E}_{i},\hat{W}_{ci}\right)\mathscr{F}_{j}\left(\mathcal{E}_{j}\right)+Q_{i}\left(e_{i}\right).\label{eq:CLNNDeltai}
\end{multline}
Approximation of the feedback-Nash equilibrium policies is realized
by tuning the estimates $\hat{V}_{i}$ and $\hat{\mu}_{i}$ so as
to minimize the BEs $\delta_{i}$. However, computation of $\delta_{i}$
in (\ref{eq:CLNNDeltai}) and $u_{ij}$ in (\ref{eq:CLNNmu_i}) requires
exact model knowledge. In the following, a CL-based system identifier
is developed to relax the exact model knowledge requirement and to
facilitate the implementation of model-based RL via BE extrapolation
(cf. \cite{Kamalapurkar.Andrews.eatoappear}). In particular, the
developed controllers do not require the knowledge of the system drift
functions $f_{i}$. 

\section{\label{sec:CLNNSysID}System Identification}

On any compact set $\chi\subset\mathbb{R}^{n}$ the function $f_{i}$
can be represented using a NN as 
\begin{equation}
f_{i}\left(x\right)=\theta_{i}^{T}\sigma_{\theta i}\left(x\right)+\epsilon_{\theta i}\left(x\right),\label{eq:CLNNfnn}
\end{equation}
for all $x\in\mathbb{R}^{n}$, where $\theta_{i}\in\mathbb{R}^{P_{i}+1\times n}$
denote the unknown output-layer NN weights, $\sigma_{\theta i}:\mathbb{R}^{n}\to\mathbb{R}^{P_{i}+1}$
denotes a bounded NN basis function, $\epsilon_{\theta i}:\mathbb{R}^{n}\to\mathbb{R}^{n}$
denotes the function reconstruction error, and $P_{i}\in\mathbb{N}$
denotes the number of NN neurons. Using the universal function approximation
property of single layer NNs, provided the rows of $\sigma_{\theta i}\left(x\right)$
form a proper basis, there exist constant ideal weights $\theta_{i}$
and positive constants $\overline{\theta_{i}}\in\mathbb{R}$ and $\overline{\epsilon_{\theta i}}\in\mathbb{R}$
such that $\left\Vert \theta_{i}\right\Vert _{F}\leq\overline{\theta_{i}}<\infty$
and $\sup_{x\in\chi}\left\Vert \epsilon_{\theta i}\left(x\right)\right\Vert \leq\overline{\epsilon_{\theta i}}$,
where $\left\Vert \cdot\right\Vert _{F}$ denotes the Frobenius norm,
i.e., $\left\Vert \theta\right\Vert _{F}\triangleq\sqrt{\mbox{tr}\left(\theta^{T}\theta\right)}$.
\begin{assumption}
\label{ass:CLNNNNKnown}The bounds $\overline{\theta_{i}}$ and $\overline{\epsilon_{\theta i}}$
are known for all $i\in\mathcal{N}.$
\end{assumption}
Using an estimate $\hat{\theta}_{i}\in\mathbb{R}^{P_{i}+1\times n}$
of the weight matrix $\theta_{i},$ the function $f_{i}$ can be approximated
by the function $\hat{f}_{i}:\mathbb{R}^{n}\times\mathbb{R}^{P_{i}+1\times n}\to\mathbb{R}^{n}$
defined by $\hat{f}_{i}\left(x,\hat{\theta}\right)\triangleq\hat{\theta}^{T}\sigma_{\theta i}\left(x\right).$
Based on (\ref{eq:CLNNfnn}), an estimator for online identification
of the drift dynamics is developed as 
\begin{equation}
\dot{\hat{x}}_{i}=\hat{\theta}_{i}^{T}\sigma_{\theta i}\left(x_{i}\right)+g_{i}\left(x_{i}\right)u_{i}+k_{i}\tilde{x}_{i},\label{eq:CLNNsysid}
\end{equation}
where $\tilde{x}_{i}\triangleq x_{i}-\hat{x}_{i}$, and $k_{i}\in\mathbb{R}$
is a positive constant learning gain. The following assumption facilitates
concurrent learning (CL)-based system identification.
\begin{assumption}
\cite{Chowdhary.Johnson2011a,Kamalapurkar.Walters.ea2016} \label{ass:CLNNsigmabar}A
history stack containing recorded state-action pairs $\left\{ x_{i}^{k},u_{i}^{k}\right\} _{k=1}^{M_{\theta i}}$
along with numerically computed state derivatives $\left\{ \dot{\bar{x}}_{i}^{k}\right\} _{k=1}^{M_{\theta i}}$
that satisfies 
\begin{gather}
\lambda_{\min}\left(\sum_{k=1}^{M_{\theta i}}\sigma_{\theta i}^{k}\left(\sigma_{\theta i}^{k}\right)^{T}\right)=\underline{\sigma_{\theta i}}>0,\nonumber \\
\left\Vert \dot{\bar{x}}_{i}^{k}-\dot{x}_{i}^{k}\right\Vert <\overline{d_{i}},\:\forall k\label{eq:CLNNsingularcond}
\end{gather}
is available a priori. In (\ref{eq:CLNNsingularcond}), $\sigma_{\theta i}^{k}\triangleq\sigma_{\theta i}\left(x_{i}^{k}\right)$,
$\overline{d_{i}},\underline{\sigma_{\theta i}}\in\mathbb{R}$ are
known positive constants, and $\lambda_{\min}\left(\cdot\right)$
denotes the minimum eigenvalue.%
\end{assumption}
The weight estimates $\hat{\theta}_{i}$ are updated using the following
CL-based update law:
\begin{equation}
\dot{\hat{\theta}}_{i}\!=\!k_{\theta i}\Gamma_{\theta i}\!\sum_{k=1}^{M_{\theta i}}\!\sigma_{\theta i}^{k}\!\left(\!\dot{\bar{x}}_{i}^{k}\!-\!g_{i}^{k}u_{i}^{k}\!-\!\hat{\theta}_{i}^{T}\sigma_{\theta i}^{k}\!\right)^{T}\!+\!\Gamma_{\theta i}\sigma_{\theta i}\!\left(x_{i}\right)\!\tilde{x}_{i}^{T},\label{eq:CLNNThetahatdot}
\end{equation}
where $g_{i}^{k}\triangleq g_{i}\left(x_{i}^{k}\right)$, $k_{\theta i}\in\mathbb{R}$
is a constant positive CL gain, and $\Gamma_{\theta i}\in\mathbb{R}^{P_{i}+1\times P_{i}+1}$
is a constant, diagonal, and positive definite adaptation gain matrix.

To facilitate the subsequent stability analysis, a candidate Lyapunov
function $V_{0i}:\mathbb{R}^{n}\times\mathbb{R}^{P_{i}+1\times n}\to\mathbb{R}$
is selected as
\begin{equation}
V_{0i}\left(\tilde{x}_{i},\tilde{\theta}_{i}\right)\triangleq\frac{1}{2}\tilde{x}_{i}^{T}\tilde{x}_{i}+\frac{1}{2}\mbox{tr}\left(\tilde{\theta}_{i}^{T}\Gamma_{\theta i}^{-1}\tilde{\theta}_{i}\right),\label{eq:CLNNV0}
\end{equation}
where $\tilde{\theta}_{i}\triangleq\theta_{i}-\hat{\theta}_{i}$ and
$\mbox{tr}\left(\cdot\right)$ denotes the trace of a matrix. Using
(\ref{eq:CLNNsysid})-(\ref{eq:CLNNThetahatdot}), the identity $\mbox{tr}\left(\tilde{\theta}^{T}\left(\sum_{j=1}^{M_{\theta i}}\sigma_{\theta i}^{j}\sigma_{\theta i}^{j}\right)\tilde{\theta}\right)=\left(\mbox{vec}\left(\tilde{\theta}_{i}\right)\right)^{T}\left(\left(\sum_{j=1}^{M_{\theta i}}\sigma_{\theta i}^{j}\sigma_{\theta i}^{j}\right)\otimes I_{p+1}\right)\left(\mbox{vec}\left(\tilde{\theta}_{i}\right)\right)$,
and the facts that $\lambda_{\min}\left\{ \left(\left(\sum_{j=1}^{M_{\theta i}}\sigma_{\theta i}^{j}\sigma_{\theta i}^{j}\right)\otimes I_{p+1}\right)\right\} $
$=$ $\lambda_{\min}\left\{ \sum_{j=1}^{M_{\theta i}}\sigma_{\theta i}^{j}\sigma_{\theta i}^{j}\right\} $
and $\lambda_{\max}\left\{ \left(\left(\sum_{j=1}^{M_{\theta i}}\sigma_{\theta i}^{j}\sigma_{\theta i}^{j}\right)\otimes I_{p+1}\right)\right\} $
$=\lambda_{\max}\left\{ \sum_{j=1}^{M_{\theta i}}\sigma_{\theta i}^{j}\sigma_{\theta i}^{j}\right\} $
(cf. \cite[Theorem 4.2.12]{Horn.Johnson1991}), the following bound
on the time derivative of $V_{0i}$ is established:%
\begin{equation}
\dot{V}_{0i}\!\leq\!-k_{i}\!\left\Vert \tilde{x}_{i}\right\Vert ^{2}\!-k_{\theta i}\underline{\sigma_{\theta i}}\!\left\Vert \tilde{\theta}_{i}\right\Vert _{F}^{2}\!+\overline{\epsilon_{\theta i}}\!\left\Vert \tilde{x}_{i}\right\Vert \!+k_{\theta i}\overline{d_{\theta i}}\!\left\Vert \tilde{\theta}_{i}\right\Vert _{F},\label{eq:CLNNV0Dot}
\end{equation}
where $\overline{d_{\theta i}}\triangleq\overline{d}_{i}\sum_{k=1}^{M_{\theta i}}\left\Vert \sigma_{\theta i}^{k}\right\Vert +\sum_{k=1}^{M_{\theta i}}\left(\left\Vert \epsilon_{\theta i}^{k}\right\Vert \left\Vert \sigma_{\theta i}^{k}\right\Vert \right)$.
Using (\ref{eq:CLNNV0}) and (\ref{eq:CLNNV0Dot}), a Lyapunov-based
stability analysis can be used to show that $\hat{\theta}_{i}$ converges
exponentially to a neighborhood around $\theta_{i}$.

\section{Approximation of the BE and the relative steady-state controller}

Using the approximations $\hat{f}_{i}$ for the functions $f_{i}$,
the BEs in (\ref{eq:CLNNDeltai}) can be approximated as
\begin{multline}
\hat{\delta}_{i}\!\left(\!\mathcal{E}_{i},\!\hat{W}_{ci},\!\left(\!\hat{W}_{a}\!\right)_{\mathcal{S}_{i}}\!,\hat{\theta}_{\mathcal{S}_{i}}\!\right)\!\triangleq\hat{\mu}_{i}^{T}\!\left(\!\mathcal{E}_{i},\hat{W}_{ai}\!\right)\!R_{i}\hat{\mu}_{i}\!\left(\!\mathcal{E}_{i},\hat{W}_{ai}\!\right)\\
+\nabla_{x_{i}}\hat{V}_{i}\left(\!\mathcal{E}_{i},\hat{W}_{ci}\!\right)\!\!\left(\!\hat{\mathcal{F}}_{i}\!\left(\!\mathcal{E}_{i},\hat{\theta}_{\mathcal{S}_{i}}\!\right)\!+\!\mathcal{G}_{i}\left(\mathcal{E}_{i}\right)\hat{\mu}_{\mathcal{S}_{i}}\!\left(\!\mathcal{E}_{i},\left(\hat{W}_{a}\right)_{\mathcal{S}_{j}}\right)\!\!\right)\\
+\sum_{j\in\mathcal{S}_{i}}\nabla_{e_{j}}\hat{V}_{i}\left(\mathcal{E}_{i},\hat{W}_{ci}\right)\mathscr{G}_{j}\left(\mathcal{E}_{j}\right)\hat{\mu}_{\mathcal{S}_{j}}\left(\mathcal{E}_{j},\left(\hat{W}_{a}\right)_{\mathcal{S}_{j}}\right)\\
+\sum_{j\in\mathcal{S}_{i}}\nabla_{e_{j}}\hat{V}_{i}\left(\mathcal{E}_{i},\hat{W}_{ci}\right)\hat{\mathscr{F}}_{j}\left(\mathcal{E}_{j},\hat{\theta}_{\mathcal{S}_{j}}\right)+Q_{i}\left(e_{i}\right).\label{eq:CLNNBE}
\end{multline}
In (\ref{eq:CLNNBE}),%
{} 
\begin{gather*}
\hat{\mathscr{F}}_{i}\left(\mathcal{E}_{i},\hat{\theta}_{\mathcal{S}_{i}}\right)\triangleq{\textstyle \sum^{i}}a_{ij}\left(\hat{f}_{i}\left(x_{i},\hat{\theta}_{i}\right)-\hat{f}_{j}\left(x_{j},\hat{\theta}_{j}\right)\right)\\
\!+\!{\textstyle \sum^{i}}\!a_{ij}\left(g_{i}\left(x_{i}\right)\mathscr{L}_{gi}^{i}\!-\!g_{j}\left(x_{j}\right)\mathscr{L}_{gi}^{j}\right)\hat{F}_{i}\left(\!\mathcal{E}_{i},\hat{\theta}_{\mathcal{S}_{i}}\!\right),\\
\hat{\mathcal{F}}_{i}\left(\mathcal{E}_{i},\hat{\theta}_{\mathcal{S}_{i}}\right)\triangleq\hat{\theta}_{i}^{T}\sigma_{\theta i}\left(x_{i}\right)+g_{i}\left(x_{i}\right)\mathscr{L}_{gi}^{i}\hat{F}_{i}\left(\mathcal{E}_{i},\hat{\theta}_{\mathcal{S}_{i}}\right),\\
\hat{F}_{i}\!\left(\!\mathcal{E}_{i},\hat{\theta}_{\mathcal{S}_{i}}\!\right)\!\triangleq\!\left[\begin{array}{c}
\left(\sum^{\lambda_{i}^{1}}a_{\lambda_{i}^{1}j}\hat{f}_{\lambda_{i}^{1}j}\!\left(x_{\lambda_{i}^{1}},\hat{\theta}_{\lambda_{i}^{1}},x_{j},\hat{\theta}_{j}\right)\!\right)\\
\vdots\\
\left(\sum^{\lambda_{i}^{s_{i}}}a_{\lambda_{i}^{s_{i}}j}\hat{f}_{\lambda_{i}^{s_{i}}j}\!\left(x_{\lambda_{i}^{s_{i}}},\hat{\theta}_{\lambda_{i}^{s_{i}}},x_{j},\hat{\theta}_{j}\right)\!\right)
\end{array}\right],\\
\hat{f}_{ij}\left(x_{i},\hat{\theta}_{i},x_{j},\hat{\theta}_{j}\right)\triangleq g_{i}^{+}\left(x_{j}+x_{dij}\right)\hat{f}_{j}\left(x_{j},\hat{\theta}_{j}\right)\\
-g_{i}^{+}\left(x_{j}+x_{dij}\right)\hat{f}_{i}\left(x_{j}+x_{dij},\hat{\theta}_{i}\right).
\end{gather*}
 The approximations $\hat{F}_{i}$, $\hat{\mathscr{F}}_{i}$, and
$\hat{\mathcal{F}}_{i}$ are related to the original unknown functions
as $\hat{F}_{i}\left(\mathcal{E}_{i},\theta_{\mathcal{S}_{i}}\right)+B_{i}\left(\mathcal{E}_{i}\right)=F_{i}\left(\mathcal{E}_{i}\right)$,
$\hat{\mathscr{F}}_{i}\left(\mathcal{E}_{i},\theta_{\mathcal{S}_{i}}\right)+\mathscr{B}_{i}\left(\mathcal{E}_{i}\right)=\mathscr{F}_{i}\left(\mathcal{E}_{i}\right)$,
and $\hat{\mathcal{F}}_{i}\left(\mathcal{E}_{i},\theta_{\mathcal{S}_{i}}\right)+\mathcal{B}_{i}\left(\mathcal{E}_{i}\right)=\mathcal{F}_{i}\left(\mathcal{E}_{i}\right)$,
where $B_{i}$, $\mathscr{B}_{i}$, and $\mathcal{B}_{i}$ are $O\left(\left(\overline{\epsilon_{\theta}}\right)_{\mathcal{S}_{i}}\right)$
terms that denote bounded function approximation errors.%
{} 

Using the approximations $\hat{f}_{i}$, an implementable form of
the controllers in (\ref{eq:CLNNU_S_i}) is expressed as
\begin{equation}
u_{\mathcal{S}_{i}}=\mathscr{L}_{gi}^{-1}\left(\mathcal{E}_{i}\right)\hat{\mu}_{\mathcal{S}_{i}}\left(\mathcal{E}_{i},\left(\hat{W}_{a}\right)_{\mathcal{S}_{i}}\right)+\mathscr{L}_{gi}^{-1}\hat{F}_{i}\left(\mathcal{E}_{i},\theta_{\mathcal{S}_{i}}\right).\label{eq:CLNNu_S_i}
\end{equation}
Using (\ref{eq:CLNNmu_S_i}) and (\ref{eq:CLNNu_S_i}), an unmeasurable
form of the virtual controllers implemented on the systems (\ref{eq:CLNNe_iDot})
and (\ref{eq:CLNNx_iDot}) is given by%
\begin{equation}
\mu_{\mathcal{S}_{i}}=\hat{\mu}_{\mathcal{S}_{i}}\left(\mathcal{E}_{i},\left(\hat{W}_{a}\right)_{\mathcal{S}_{i}}\right)-\hat{F}_{i}\left(\mathcal{E}_{i},\tilde{\theta}_{\mathcal{S}_{i}}\right)-B_{i}\left(\mathcal{E}_{i}\right).\label{eq:CLNNmu_S_iUnm}
\end{equation}

\section{Value function approximation}

On any compact set $\chi\in\mathbb{R}^{n\left(s_{i}+1\right)}$, the
value functions can be represented as
\begin{equation}
V_{i}^{*}\left(\mathcal{E}_{i}\right)=W_{i}^{T}\sigma_{i}\left(\mathcal{E}_{i}\right)+\epsilon_{i}\left(\mathcal{E}_{i}\right),\:\forall\mathcal{E}_{i}\in\mathbb{R}^{n\left(s_{i}+1\right)},\label{eq:CLNNV_i*NN}
\end{equation}
where $W_{i}\in\mathbb{R}^{L_{i}}$ are ideal NN weights, $\sigma_{i}:\mathbb{R}^{n\left(s_{i}+1\right)}\to\mathbb{R}^{L_{i}}$
are NN basis functions and $\epsilon_{i}:\mathbb{R}^{n\left(s_{i}+1\right)}\to\mathbb{R}$
are function approximation errors. Using the universal function approximation
property of single layer NNs, provided $\sigma_{i}\left(\mathcal{E}_{i}\right)$
forms a proper basis, there exist constant ideal weights $W_{i}$
and positive constants $\overline{W_{i}}\in\mathbb{R}$ and $\overline{\epsilon_{i}},\overline{\nabla\epsilon_{i}}\in\mathbb{R}$
such that $\left\Vert W_{i}\right\Vert \leq\overline{W_{i}}<\infty$,
$\sup_{\mathcal{E}_{i}\in\chi}\left\Vert \epsilon_{i}\left(\mathcal{E}_{i}\right)\right\Vert \leq\overline{\epsilon_{i}}$,
and $\sup_{\mathcal{E}_{i}\in\chi}\left\Vert \nabla\epsilon_{i}\left(\mathcal{E}_{i}\right)\right\Vert \leq\overline{\nabla\epsilon_{i}}$. 
\begin{assumption}
\label{ass:CLNNVFNNKnown} The constants $\overline{\epsilon_{i}},$
$\overline{\nabla\epsilon_{i}},$ and $\overline{W_{i}}$ are known
for all $i\in\mathcal{N}$.
\end{assumption}
Using (\ref{eq:CLNNmu_i*}) and (\ref{eq:CLNNV_i*NN}), the feedback-Nash
equilibrium policies are%
\[
\mu_{i}^{*}\left(\mathcal{E}_{i}\right)=-\frac{1}{2}R_{i}^{-1}G_{\sigma i}\left(\mathcal{E}_{i}\right)W_{i}-\frac{1}{2}R_{i}^{-1}G_{\epsilon i}\left(\mathcal{E}_{i}\right),
\]
for all $\mathcal{E}_{i}\in\mathbb{R}^{n\left(s_{i}+1\right)},$ where
$G_{\sigma i}\left(\mathcal{E}_{i}\right)\triangleq\sum_{j\in\mathcal{S}_{i}}\left(\mathscr{G}_{j}^{i}\left(\mathcal{E}_{i}\right)\right)^{T}\left(\nabla_{e_{j}}\sigma_{i}\left(\mathcal{E}_{i}\right)\right)^{T}+\left(\mathcal{G}_{i}^{i}\left(\mathcal{E}_{i}\right)\right)^{T}\left(\nabla_{x_{i}}\sigma_{i}\left(\mathcal{E}_{i}\right)\right)^{T}$
and $G_{\epsilon i}\left(\mathcal{E}_{i}\right)\triangleq\sum_{j\in\mathcal{S}_{i}}\left(\mathscr{G}_{j}^{i}\left(\mathcal{E}_{i}\right)\right)^{T}\left(\nabla_{e_{j}}\epsilon_{i}\left(\mathcal{E}_{i}\right)\right)^{T}+\left(\mathcal{G}_{i}^{i}\left(\mathcal{E}_{i}\right)\right)^{T}\left(\nabla_{x_{i}}\epsilon_{i}\left(\mathcal{E}_{i}\right)\right)^{T}.$
The value functions and the policies are approximated using NNs as
\begin{align}
\hat{V}_{i}\left(\mathcal{E}_{i},\hat{W}_{ci}\right) & \triangleq\hat{W}_{ci}^{T}\sigma_{i}\left(\mathcal{E}_{i}\right),\nonumber \\
\hat{\mu}_{i}\left(\mathcal{E}_{i},\hat{W}_{ai}\right) & \triangleq-\frac{1}{2}R_{i}^{-1}G_{\sigma i}\left(\mathcal{E}_{i}\right)\hat{W}_{ai},\label{eq:CLNNmuhatVhat}
\end{align}
where $\hat{W}_{ci}$ and $\hat{W}_{ai}$ are estimates of the ideal
weights $W_{i}$, introduced in (\ref{eq:CLNNDeltai}).

\section{Simulation of experience via BE extrapolation}

A consequence of Theorem \ref{thm:CLNNNesSuf} is that the BE provides
an indirect measure of how close the estimates $\hat{W}_{ci}$ and
$\hat{W}_{ai}$ are to the ideal weights $W_{i}$. From a reinforcement
learning perspective, each evaluation of the BE along the system trajectory
can be interpreted as experience gained by the critic, and each evaluation
of the BE at points not yet visited can be interpreted as simulated
experience. In previous results such as \cite{Vamvoudakis2011,Johnson2011a,Vamvoudakis.Lewis.ea2012a,Kamalapurkar.Dinh.ea2013,Modares.Lewis.ea2014},
the critic is restricted to the experience gained (in other words
BEs evaluated) along the system state trajectory. The development
in \cite{Vamvoudakis2011,Johnson2011a,Vamvoudakis.Lewis.ea2012a,Kamalapurkar.Dinh.ea2013}
can be extended to employ simulated experience; however, the extension
requires exact model knowledge. In results such as \cite{Modares.Lewis.ea2014},
the formulation of the BE does not allow for simulation of experience.
The formulation in (\ref{eq:CLNNBE}) employs the system identifier
developed in Section \ref{sec:CLNNSysID} to facilitate approximate
evaluation of the BE at off-trajectory points.

To simulate experience, a set of points $\left\{ \mathcal{E}_{i}^{k}\right\} _{k=1}^{M_{i}}$
is selected corresponding to each agent $i$ , and the instantaneous
BE in (\ref{eq:CLNNDeltai}) is approximated at the current state
and at the selected points using (\ref{eq:CLNNdeltahat_i}). The approximation
at the current state is denoted by $\hat{\delta}_{ti}$ and the approximation
at the selected points is denoted by $\hat{\delta}_{ti}^{k}$, where
$\hat{\delta}_{ti}$ and $\hat{\delta}_{ti}^{k}$ are defined as 
\begin{align*}
\hat{\delta}_{ti}\left(t\right) & \triangleq\hat{\delta}_{i}\left(\mathcal{E}_{i}\left(t\right),\hat{W}_{ci}\left(t\right),\left(\hat{W}_{a}\left(t\right)\right)_{\mathcal{S}_{i}},\left(\hat{\theta}\left(t\right)\right)_{\mathcal{S}_{i}}\right),\\
\hat{\delta}_{ti}^{k}\left(t\right) & \triangleq\hat{\delta}_{i}\left(\mathcal{E}_{i}^{k},\hat{W}_{ci}\left(t\right),\left(\hat{W}_{a}\left(t\right)\right)_{\mathcal{S}_{i}},\left(\hat{\theta}\left(t\right)\right)_{\mathcal{S}_{i}}\right).
\end{align*}
Note that once $\left\{ e_{j}\right\} _{j\in\mathcal{S}_{i}}$ and
$x_{i}$ are selected, the $i\textsuperscript{th}$ agent can compute
the states of all the remaining agents in the sub-graph. For notational
brevity, the arguments to the functions $\sigma_{i}$, $\hat{\mathscr{F}}_{i}$,
$\mathscr{G}_{i}$, $\mathcal{G}_{i}$, $\hat{\mathcal{F}}_{i}$,
$\hat{\mu}_{i}$, $G_{\sigma i}$, $G_{\epsilon i}$, and $\epsilon_{i}$
are suppressed hereafter.

The critic uses simulated experience to update the value function
weights using a least squares-based update law
\begin{gather}
\dot{\hat{W}}_{ci}=-\eta_{c1i}\Gamma_{i}\frac{\omega_{i}}{\rho_{i}}\hat{\delta}_{ti}-\frac{\eta_{c2i}\Gamma_{i}}{M_{i}}\sum_{k=1}^{M_{i}}\frac{\omega_{i}^{k}}{\rho_{i}^{k}}\hat{\delta}_{ti}^{k},\nonumber \\
\dot{\Gamma}_{i}\!=\!\left(\!\beta_{i}\Gamma_{i}-\eta_{c1i}\Gamma_{i}\frac{\omega_{i}\omega_{i}^{T}}{\rho_{i}^{2}}\Gamma_{i}\!\right)\!\mathbf{1}_{\left\{ \left\Vert \Gamma_{i}\right\Vert \leq\overline{\Gamma}_{i}\right\} },\label{eq:CLNNCriticUpdate}
\end{gather}
where $\rho_{i}\triangleq1+\nu_{i}\omega_{i}^{T}\Gamma_{i}\omega_{i}$,
$\Gamma_{i}\in\mathbb{R}^{L_{i}\times L_{i}}$ denotes the time-varying
least-squares learning gain, $\overline{\Gamma}_{i}\in\mathbb{R}$
denotes the saturation constant, $\left\Vert \Gamma_{i}\left(t_{0}\right)\right\Vert \leq\overline{\Gamma}_{i},$
and $\eta_{c1i},\eta_{c2i},\beta_{i},\nu_{i}\in\mathbb{R}$ are constant
positive learning gains. In (\ref{eq:CLNNCriticUpdate}), %
{} %
{} 
\begin{align*}
\omega_{i} & \triangleq\sum_{j\in\mathcal{S}_{i}}\nabla_{e_{j}}\sigma_{i}\left(\hat{\mathscr{F}}_{j}+\mathscr{G}_{j}\hat{\mu}_{\mathcal{S}_{j}}\right)+\nabla_{x_{i}}\sigma_{i}\left(\hat{\mathcal{F}}_{i}+\mathcal{G}_{i}\hat{\mu}_{\mathcal{S}_{i}}\right),\\
\omega_{i}^{k} & \!\triangleq\!\sum_{j\in\mathcal{S}_{i}}\!\nabla_{e_{j}}\sigma_{i}^{k}\left(\hat{\mathscr{F}}_{j}^{k}+\mathscr{G}_{j}^{k}\hat{\mu}_{\mathcal{S}_{j}}^{k}\right)\!+\!\nabla_{x_{i}}\sigma_{i}^{k}\left(\hat{\mathcal{F}}_{i}^{k}+\mathcal{G}_{i}^{k}\hat{\mu}_{\mathcal{S}_{i}}^{k}\right),
\end{align*}
 where for a function $\phi_{i}\left(\mathcal{E}_{i},\left(\cdot\right)\right)$,
the notation $\phi_{i}^{k}$ indicates evaluation at $\mathcal{E}_{i}=\mathcal{E}_{i}^{k}$;
i.e., $\phi_{i}^{k}\triangleq\phi_{i}\left(\mathcal{E}_{i}^{k},\left(\cdot\right)\right)$.
The actor updates the policy weights using the following update law
derived based on the Lyapunov-based stability analysis in section
\ref{sec:Stability-analysis}.
\begin{multline}
\dot{\hat{W}}_{ai}=-\eta_{a2i}\hat{W}_{ai}+\frac{1}{4}\eta_{c1i}G_{\sigma i}^{T}R_{i}^{-1}G_{\sigma i}\hat{W}_{ai}\frac{\omega_{i}^{T}}{\rho_{i}}\hat{W}_{ci}\\
+\frac{1}{4}\sum_{k=1}^{M_{i}}\frac{\eta_{c2i}}{M_{i}}\left(G_{\sigma i}^{k}\right)^{T}R_{i}^{-1}G_{\sigma i}^{k}\hat{W}_{ai}\frac{\left(\omega_{i}^{k}\right)^{T}}{\rho_{i}^{k}}\hat{W}_{ci}\\
-\eta_{a1i}\left(\hat{W}_{ai}-\hat{W}_{ci}\right),\label{eq:CLNNActor update}
\end{multline}
where $\eta_{a1i},\eta_{a2i}\in\mathbb{R}$ are constant positive
learning gains. The following assumption facilitates simulation of
experience.
\begin{assumption}
\label{ass:CLNNCLW}\cite{Kamalapurkar.Walters.ea2016} For each $i\in\mathcal{N}$,
there exists a finite set of points $\left\{ \mathcal{E}_{i}^{k}\right\} _{k=1}^{M_{i}}$
such that 
\begin{gather}
\underline{\rho_{i}}\triangleq\frac{\left(\inf_{t\in\mathbb{R}_{\geq0}}\left(\lambda_{\min}\left\{ \sum_{k=1}^{M_{i}}\frac{\omega_{i}^{k}\left(t\right)\left(\omega_{i}^{k}\right)^{T}\left(t\right)}{\rho_{i}^{k}\left(t\right)}\right\} \right)\right)}{M_{i}}>0,\label{eq:CLNNCLRank2}
\end{gather}
where $\lambda_{\min}$ denotes the minimum eigenvalue, and $\underline{\rho_{i}}\in\mathbb{R}$
is a positive constant.
\end{assumption}

\section{\label{sec:Stability-analysis}Stability analysis}

To facilitate the stability analysis, the left hand side of (\ref{eq:CLNNHJB})
is subtracted from (\ref{eq:CLNNBE}) to express the BEs in terms
of the weight estimation errors as%
\begin{multline}
\hat{\delta}_{ti}=-\tilde{W}_{ci}^{T}\omega_{i}-W_{i}^{T}\nabla_{x_{i}}\sigma_{i}\left(\mathcal{E}_{i}\right)\hat{\mathcal{F}}_{i}\left(\mathcal{E}_{i},\tilde{\theta}_{\mathcal{S}_{i}}\right)\\
+\frac{1}{4}\tilde{W}_{ai}^{T}G_{\sigma i}^{T}R_{i}^{-1}G_{\sigma i}\tilde{W}_{ai}-\frac{1}{2}W_{i}^{T}G_{\sigma i}^{T}R_{i}^{-1}G_{\sigma i}\tilde{W}_{ai}\\
+\frac{1}{2}W_{i}^{T}\sum_{j\in\mathcal{S}_{i}}\nabla_{e_{j}}\sigma_{i}\left(\mathcal{E}_{i}\right)\mathscr{G}_{j}\mathcal{R}_{\mathcal{S}_{j}}\left(\tilde{W}_{a}\right)_{\mathcal{S}_{j}}\\
-W_{i}^{T}\sum_{j\in\mathcal{S}_{i}}\nabla_{e_{j}}\sigma_{i}\left(\mathcal{E}_{i}\right)\hat{\mathscr{F}}_{j}\left(\mathcal{E}_{j},\tilde{\theta}_{\mathcal{S}_{j}}\right)\\
+\frac{1}{2}W_{i}^{T}\nabla_{x_{i}}\sigma_{i}\left(\mathcal{E}_{i}\right)\mathcal{G}_{i}\mathcal{R}_{\mathcal{S}_{i}}\left(\tilde{W}_{a}\right)_{\mathcal{S}_{i}}+\Delta_{i},\label{eq:CLNNdeltahat_i}
\end{multline}
where $\tilde{\left(\cdot\right)}\triangleq\left(\cdot\right)-\hat{\left(\cdot\right)}$,
$\Delta_{i}=O\left(\left(\overline{\epsilon}\right)_{\mathcal{S}_{i}},\left(\overline{\nabla\epsilon}\right)_{\mathcal{S}_{i}},\left(\overline{\epsilon_{\theta}}\right)_{\mathcal{S}_{i}}\right)$,
and $\mathcal{R}_{\mathcal{S}_{j}}\triangleq\mbox{diag}\left(\left[R_{\lambda_{j}^{1}}^{-1}G_{\sigma\lambda_{j}^{1}}^{T},\cdots,R_{\lambda_{j}^{s_{j}}}^{-1}G_{\sigma\lambda_{j}^{s_{j}}}^{T}\right]\right)$
are block diagonal matrices. Consider a set of extended neighbors
$\mathcal{S}_{p}$ corresponding to the $p$\textsuperscript{th}
agent. To analyze asymptotic properties of the agents in $\mathcal{S}_{p},$
consider the following candidate Lyapunov function 
\begin{multline}
V_{Lp}\left(Z_{p},t\right)\triangleq\sum_{i\in\mathcal{S}_{p}}V_{ti}\left(e_{\mathcal{S}_{i}},t\right)+\sum_{i\in\mathcal{S}_{p}}\frac{1}{2}\tilde{W}_{ci}^{T}\Gamma_{i}^{-1}\tilde{W}_{ci}\\
+\sum_{i\in\mathcal{S}_{p}}\frac{1}{2}\tilde{W}_{ai}^{T}\tilde{W}_{ai}+\sum_{i\in\mathcal{S}_{p}}V_{0i}\left(\tilde{x}_{i},\tilde{\theta}_{i}\right),\label{eq:CLNNV_Lp}
\end{multline}
where $Z_{p}\in\mathbb{R}^{\left(2ns_{i}+2L_{i}s_{i}+n\left(P_{i}+1\right)s_{i}\right)}$
is defined as 
\[
Z_{p}\triangleq\left[e_{\mathcal{S}_{p}}^{T},\left(\tilde{W}_{c}\right)_{\mathcal{S}_{p}}^{T},\left(\tilde{W}_{a}\right)_{\mathcal{S}_{p}}^{T},\tilde{x}_{\mathcal{S}_{p}}^{T},\mbox{vec}\left(\tilde{\theta}_{\mathcal{S}_{p}}\right)^{T}\right]^{T},
\]
$\mbox{vec}\left(\cdot\right)$ denotes the vectorization operator,
and $V_{ti}:\mathbb{R}^{ns_{i}}\times\mathbb{R}\to\mathbb{R}$ is
defined as 
\begin{equation}
V_{ti}\left(e_{\mathcal{S}_{i}},t\right)\triangleq V_{i}^{*}\left(\left[e_{\mathcal{S}_{i}}^{T},\:x_{i}^{T}\left(t\right)\right]^{T}\right),\label{eq:CLNNV_ti}
\end{equation}
for all $e_{\mathcal{S}_{i}}\in\mathbb{R}^{ns_{i}}$ and for all $t\in\mathbb{R}_{\geq t_{0}}$.
Since $V_{ti}^{*}$ depends on $t$ only through uniformly bounded
leader trajectories%
, Lemma 1 from \cite{Kamalapurkar.Dinh.ea2015} can be used to show
that $V_{ti}$ is a positive definite and decrescent function.\footnote{Since the graph has a spanning tree, the mapping between the errors
and the states is invertible. Hence, the state of an agent can be
expressed as $x_{i}=h_{i}\left(e_{\mathcal{S}_{i}},x_{0}\right)$
for some function $h_{i}$. Thus, the value function can be expressed
as $V_{i}^{*}\left(e_{\mathcal{S}_{i}},x_{0}\right)=V_{i}^{*}\left(e_{\mathcal{S}_{i}},h\left(e_{\mathcal{S}_{i}},x_{0}\right)\right)$.
Then, $V_{ti}^{*}$ can be alternatively defined as $V_{ti}\left(e_{\mathcal{S}_{i}},t\right)\triangleq V_{i}^{*}\left(\begin{bmatrix}e_{\mathcal{S}_{i}}\\
x_{0}\left(t\right)
\end{bmatrix}\right).$ Since $x_{0}$ is a uniformly bounded function of $t$ by assumption,
Lemma 1 from \cite{Kamalapurkar.Dinh.ea2015} can be used to conclude
that $V_{ti}$ is a positive definite and decrescent function.} Thus, using Lemma 4.3 from \cite{Khalil2002}, the following bounds
on the candidate Lyapunov function in (\ref{eq:CLNNV_Lp}) are established
\begin{equation}
\underline{v_{lp}}\left(\left\Vert Z_{p}\right\Vert \right)\leq V_{Lp}\left(Z_{p},t\right)\leq\overline{v_{lp}}\left(\left\Vert Z_{p}\right\Vert \right),\label{eq:CLNNVbounds}
\end{equation}
 for all $Z_{p}\in\mathbb{R}^{\left(2ns_{i}+2L_{i}s_{i}+n\left(P_{i}+1\right)s_{i}\right)}$
and for all $t\in\mathbb{R}_{\geq t_{0}}$, where $\underline{v_{lp}},\overline{v_{lp}}:\mathbb{R}\to\mathbb{R}$
are class $\mathcal{K}$ functions. 

To facilitate the stability analysis, given any compact ball $\chi_{p}\subset\mathbb{R}^{2ns_{i}+2L_{i}s_{i}+n\left(P_{i}+1\right)s_{i}}$
of radius $r_{p}\in\mathbb{R}$ centered at the origin, a positive
constant $\iota_{p}\in\mathbb{R}$ is defined as
\begin{gather*}
\iota_{p}\triangleq\sum_{i\in\mathcal{S}_{p}}\overline{\left\Vert \sum_{j\in\mathcal{S}_{i}}\nabla_{e_{j}}V_{i}^{*}\left(\mathcal{E}_{i}\right)\mathscr{G}_{j}B_{j}+\nabla_{x_{i}}V_{i}^{*}\left(\mathcal{E}_{i}\right)\mathcal{G}_{i}B_{i}\right\Vert }\\
+\!\!\sum_{i\in\mathcal{S}_{p}}\!\!\frac{1}{2}\text{\ensuremath{\overline{\left\Vert \nabla_{x_{i}}V_{i}^{*}\left(\mathcal{E}_{i}\right)\mathcal{G}_{i}\mathcal{R}_{\mathcal{S}_{i}}\epsilon_{\mathcal{S}_{i}}\!\!+\!\!\sum_{j\in\mathcal{S}_{i}}\nabla_{e_{j}}V_{i}^{*}\left(\mathcal{E}_{i}\right)\mathscr{G}_{j}\mathcal{R}_{\mathcal{S}_{j}}\epsilon_{\mathcal{S}_{j}}\right\Vert }}}\\
+\sum_{i\in\mathcal{S}_{p}}\frac{\overline{\epsilon_{\theta i}}^{2}}{2k_{i}}+\sum_{i\in\mathcal{S}_{p}}\frac{3\left(k_{\theta i}\overline{d_{\theta i}}+\overline{\left\Vert A_{i}^{\theta}\right\Vert }\overline{\left\Vert B_{i}^{\theta}\right\Vert }\right)^{2}}{4\underline{\sigma_{\theta i}}}\\
+\sum_{i\in\mathcal{S}_{p}}\frac{3}{4\left(\eta_{a1i}+\eta_{a2i}\right)}\Biggl(\frac{1}{2}\overline{\left\Vert A_{i}^{a1}\right\Vert }+\eta_{a2i}\overline{W_{i}}\\
+\frac{1}{4}\left(\eta_{c1i}+\eta_{c2i}\right)\overline{\left\Vert W_{i}^{T}\frac{\omega_{i}}{\rho_{i}}W_{i}^{T}G_{\sigma i}^{T}R_{i}^{-1}G_{\sigma i}\right\Vert }\Biggl)^{2}\\
+\sum_{i\in\mathcal{S}_{p}}\frac{5\left(\eta_{c1i}+\eta_{c2i}\right)^{2}\overline{\left\Vert \frac{\omega_{i}}{\rho_{i}}\Delta_{i}\right\Vert }^{2}}{4\eta_{c2i}\underline{\rho_{i}}}
\end{gather*}
where for any function $\varpi:\mathbb{R}^{l}\to\mathbb{R}$, $l\in\mathbb{N}$,
the notation $\overline{\left\Vert \varpi\right\Vert }$ denotes $\sup_{y\in\mathcal{\chi}_{p}\cap\mathbb{R}^{l}}\left\Vert \varpi\left(y\right)\right\Vert $
and $A_{i}^{\theta}$, $B_{i}^{\theta}$, and $A_{i}^{a1}$ are uniformly
bounded state-dependent terms. The following sufficient gain conditions
facilitate the subsequent stability analysis. 
\begin{equation}
\frac{\eta_{c2i}\underline{\rho_{i}}}{5}\!\!>\!\!\sum_{j\in\mathcal{S}_{p}}\!\!\frac{3s_{p}\mathbf{1}_{j\in\mathcal{S}_{i}}\left(\eta_{c1i}+\eta_{c2i}\right)^{2}\overline{\left\Vert A_{ij}^{1a\theta}\right\Vert }^{2}\overline{\left\Vert B_{ij}^{1a\theta}\right\Vert }^{2}}{4k_{\theta j}\underline{\sigma_{\theta j}}},\label{eq:CLNNSuffCond1}
\end{equation}
\begin{gather}
\frac{\left(\eta_{a1i}+\eta_{a2i}\right)}{3}>\sum_{j\in\mathcal{S}_{p}}\frac{5s_{p}\mathbf{1}_{i\in\mathcal{S}_{j}}\left(\eta_{c1j}+\eta_{c2j}\right)^{2}\overline{\left\Vert A_{ji}^{1ac}\right\Vert }^{2}}{16\eta_{c2j}\underline{\rho_{j}}}\nonumber \\
+\frac{5\eta_{a1i}^{2}}{4\eta_{c2i}\underline{\rho_{i}}}+\frac{\left(\eta_{c1i}+\eta_{c2i}\right)\overline{W_{i}}\left\Vert \frac{\omega_{i}}{\rho_{i}}\right\Vert \overline{\left\Vert G_{\sigma i}^{T}R_{i}^{-1}G_{\sigma i}\right\Vert }}{4},\label{eq:CLNNSuffCond2}
\end{gather}
\begin{equation}
v_{lp}^{-1}\left(\iota_{p}\right)<\overline{v_{lp}}^{-1}\left(\underline{v_{lp}}\left(r_{p}\right)\right),\label{eq:CLNNSuffCond3}
\end{equation}
where $A_{ij}^{1a\theta}$, $B_{ij}^{1a\theta}$, and $A_{ji}^{1ac}$
are uniformly bounded state-dependent terms.
\begin{thm}
\label{thm:CLNNMain}Provided Assumptions \ref{ass:CLNNLeader} -
\ref{ass:CLNNCLW} hold and the sufficient gain conditions in (\ref{eq:CLNNSuffCond1})-(\ref{eq:CLNNSuffCond3})
are satisfied, the controller in (\ref{eq:CLNNmuhatVhat}) along with
the actor and critic update laws in (\ref{eq:CLNNCriticUpdate}) and
(\ref{eq:CLNNActor update}), and the system identifier in (\ref{eq:CLNNsysid})
along with the weight update laws in (\ref{eq:CLNNThetahatdot}) ensure
that the local neighborhood tracking errors $e_{i}$ are ultimately
bounded and that the policies $\hat{\mu}_{i}$ converge to a neighborhood
around the feedback-Nash policies $\mu_{i}^{*}$ for all $i\in\mathcal{N}$.
\end{thm}
\begin{IEEEproof}
The time derivative of the candidate Lyapunov function in (\ref{eq:CLNNV_Lp})
is given by 
\begin{multline}
\dot{V}_{Lp}=\sum_{i\in\mathcal{S}_{p}}\dot{V}_{ti}\left(e_{\mathcal{S}_{i}},t\right)-\frac{1}{2}\sum_{i\in\mathcal{S}_{p}}\tilde{W}_{ci}^{T}\Gamma_{i}^{-1}\dot{\Gamma}_{i}\Gamma_{i}^{-1}\tilde{W}_{ci}\\
-\sum_{i\in\mathcal{S}_{p}}\tilde{W}_{ci}^{T}\Gamma_{i}^{-1}\dot{\hat{W}}_{ci}-\sum_{i\in\mathcal{S}_{p}}\tilde{W}_{ai}^{T}\dot{\hat{W}}_{ai}+\sum_{i\in\mathcal{S}_{p}}\dot{V}_{0i}\left(\tilde{x}_{i},\tilde{\theta}_{i}\right).\label{eq:CLNNVdot1}
\end{multline}
Using (\ref{eq:CLNNHJB}), (\ref{eq:CLNNV0Dot}), (\ref{eq:CLNNmu_S_iUnm}),
and (\ref{eq:CLNNdeltahat_i}), the update laws in (\ref{eq:CLNNCriticUpdate})
and (\ref{eq:CLNNActor update}), and the definition of $V_{ti}$
in (\ref{eq:CLNNV_ti}), the derivative in (\ref{eq:CLNNVdot1}) can
be bounded as\footnote{For a detailed derivation of the bound, see \cite{Kamalapurkar2014}.}%
\begin{align*}
\dot{V}_{Lp} & \leq\sum_{i\in\mathcal{S}_{p}}\left(-\frac{\eta_{c2i}\underline{\rho_{i}}}{5}\left\Vert \tilde{W}_{ci}\right\Vert ^{2}-\frac{\left(\eta_{a1i}+\eta_{a2i}\right)}{3}\left\Vert \tilde{W}_{ai}\right\Vert ^{2}\right)\\
 & +\sum_{i\in\mathcal{S}_{p}}\left(-\underline{q_{i}}\left(\left\Vert e_{i}\right\Vert \right)-\frac{k_{i}}{2}\left\Vert \tilde{x}_{i}\right\Vert ^{2}-\frac{k_{\theta i}\underline{\sigma_{\theta i}}}{3}\left\Vert \tilde{\theta}_{i}\right\Vert _{F}^{2}\right)+\iota_{p}.
\end{align*}
Let $v_{lp}:\mathbb{R}\to\mathbb{R}$ be a class $\mathcal{K}$ function
such that 
\begin{multline}
v_{lp}\left(\left\Vert Z_{p}\right\Vert \right)\leq\frac{1}{2}\sum_{i\in\mathcal{S}_{p}}\underline{q_{i}}\left(\left\Vert e_{i}\right\Vert \right)+\frac{1}{2}\sum_{i\in\mathcal{S}_{p}}\frac{\eta_{c2i}\underline{\rho_{i}}}{5}\left\Vert \tilde{W}_{ci}\right\Vert ^{2}\\
+\frac{1}{2}\sum_{i\in\mathcal{S}_{p}}\frac{\left(\eta_{a1i}+\eta_{a2i}\right)}{3}\left\Vert \tilde{W}_{ai}\right\Vert ^{2}+\frac{1}{2}\sum_{i\in\mathcal{S}_{p}}\frac{k_{i}}{2}\left\Vert \tilde{x}_{i}\right\Vert ^{2}\\
+\frac{1}{2}\sum_{i\in\mathcal{S}_{p}}\frac{k_{\theta i}\underline{\sigma_{\theta i}}}{3}\left\Vert \tilde{\theta}_{i}\right\Vert _{F}^{2},\label{eq:CLNNvlp}
\end{multline}
where $\underline{q_{i}}:\mathbb{R}\to\mathbb{R}$ are class $\mathcal{K}$
functions such that $\underline{q_{i}}\left(\left\Vert e\right\Vert \right)\leq Q_{i}\left(e\right),\:\forall e\in\mathbb{R}^{n},\:\forall i\in\mathcal{N}$.
Then, the Lyapunov derivative can be bounded as 
\begin{equation}
\dot{V}_{Lp}\leq-v_{lp}\left(\left\Vert Z_{p}\right\Vert \right)\label{eq:CLNNVdot4}
\end{equation}
for all $Z_{p}$ such that $Z_{p}\in\chi_{p}$ and $\left\Vert Z_{p}\right\Vert \geq v_{lp}^{-1}\left(\iota_{p}\right)$.
Using the bounds in (\ref{eq:CLNNVbounds}), the sufficient conditions
in (\ref{eq:CLNNSuffCond1})-(\ref{eq:CLNNSuffCond3}), and the inequality
in (\ref{eq:CLNNVdot4}), Theorem 4.18 in \cite{Khalil2002} can be
invoked to conclude that every trajectory $Z_{p}\left(t\right)$ satisfying
$\left\Vert Z_{p}\left(t_{0}\right)\right\Vert \leq\overline{v_{lp}}^{-1}\left(\underline{v_{lp}}\left(r_{p}\right)\right)$,
is bounded for all $t\in\mathbb{R}_{\geq t_{0}}$ and satisfies 
\[
\lim\sup_{t\to\infty}\left\Vert Z_{p}\left(t\right)\right\Vert \leq\underline{v_{lp}}^{-1}\left(\overline{v_{lp}}\left(v_{lp}^{-1}\left(\iota_{p}\right)\right)\right).
\]

Since the choice of the subgraph $\mathcal{S}_{p}$ was arbitrary,
the neighborhood tracking errors $e_{i}$ are ultimately bounded for
all $i\in\mathcal{N}$. Furthermore, the weight estimates $\hat{W}_{ai}$
converge to a neighborhood of the ideal weights $W_{i}$; hence, invoking
Theorem \ref{thm:CLNNNesSuf}, the policies $\hat{\mu}_{i}$ converge
to a neighborhood of the feedback-Nash equilibrium policies $\mu_{i}^{*}$
for all $i\in\mathcal{N}$.
\end{IEEEproof}

\section{Simulations}

\begin{figure}
\begin{centering}
\includegraphics[width=0.8\columnwidth]{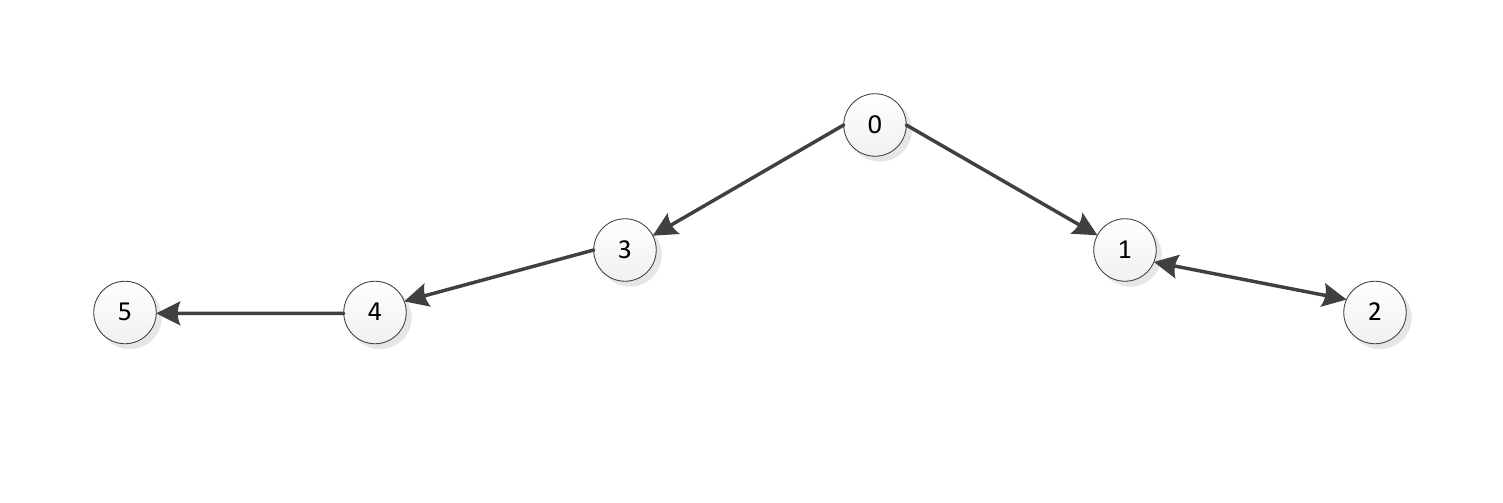} 
\par\end{centering}
\caption{\label{fig:CLNNTOPOLOGY}Communication topology: A network containing
five agents.}
\end{figure}

This section provides a simulation example to demonstrate the applicability
of the developed technique. The agents are assumed to have the communication
topology as shown in Figure \ref{fig:CLNNTOPOLOGY} with unit pinning
gains and edge weights. The motion of the agents is described by identical
nonlinear one-dimensional dynamics of the form (\ref{eq:CLNNDyn})
where $f_{i}\left(x_{i}\right)=\theta_{i1}x_{i}+\theta_{i2}x_{i}^{2}$,
and $g_{i}\left(x_{i}\right)=\left(\cos(2x_{i1})+2\right)$ for all
$i=1,\cdots,5.$ The ideal values of the unknown parameters are selected
to be $\theta_{i1}=0,$ $0,$ $0.1,$ $0.5,$ and $0.2$, and $\theta_{i2}=1$,
$0.5,$ $1,$ $1,$ and $1,$ for $i=1,\cdots,5$, respectively. The
agents start at $x_{i}=2$ for all $i$, and their final desired locations
with respect to each other are given by $x_{d12}=0.5,$ $x_{d21}=-0.5,$
$x_{d43}=-0.5,$ and $x_{d53}=-0.5$. The leader traverses an exponentially
decaying trajectory $x_{0}\left(t\right)=e^{-0.1t}$. The desired
positions of agents 1 and 3 with respect to the leader are $x_{d10}=0.75$
and $x_{d30}=1$, respectively.\footnote{The optimal control problem parameters, basis functions, and adaptation
gains for all the agents and the plots for weight estimates corresponding
to agents 1-5 are available in \cite{Kamalapurkar2014}}
\begin{figure}
\begin{centering}
\includegraphics[width=0.7\columnwidth]{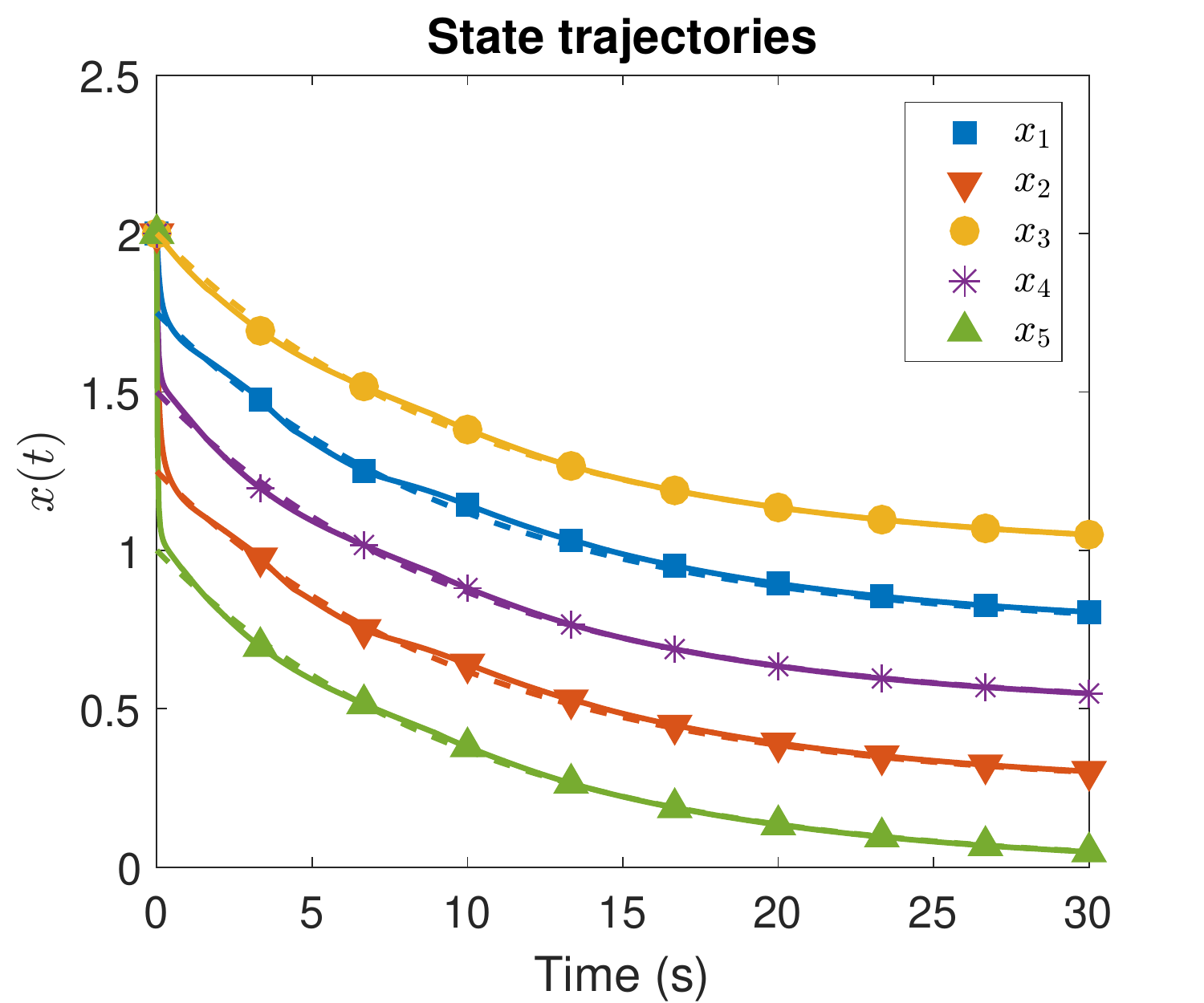} 
\par\end{centering}
\caption{State trajectories for the five agents for the one-dimensional example.
The dotted lines show the desired state trajectories.\label{fig:CLNNX}}
\end{figure}
For each agent $i,$ five values of $e_{i}$, three values of $x_{i}$,
and three values of errors corresponding to all the extended neighbors
are selected for BE extrapolation, resulting in $5\times3^{s_{i}}$
total values of $\mathcal{E}_{i}$. All agents estimate the unknown
drift parameters using history stacks containing thirty points recorded
online using a singular value maximizing algorithm (cf. \cite{Chowdhary.Yucelen.ea2012}),
and compute the required state derivatives using a fifth order Savitzky-Golay
smoothing filter (cf. \cite{Savitzky.Golay1964}).
\begin{figure}
\begin{centering}
\includegraphics[width=0.7\columnwidth]{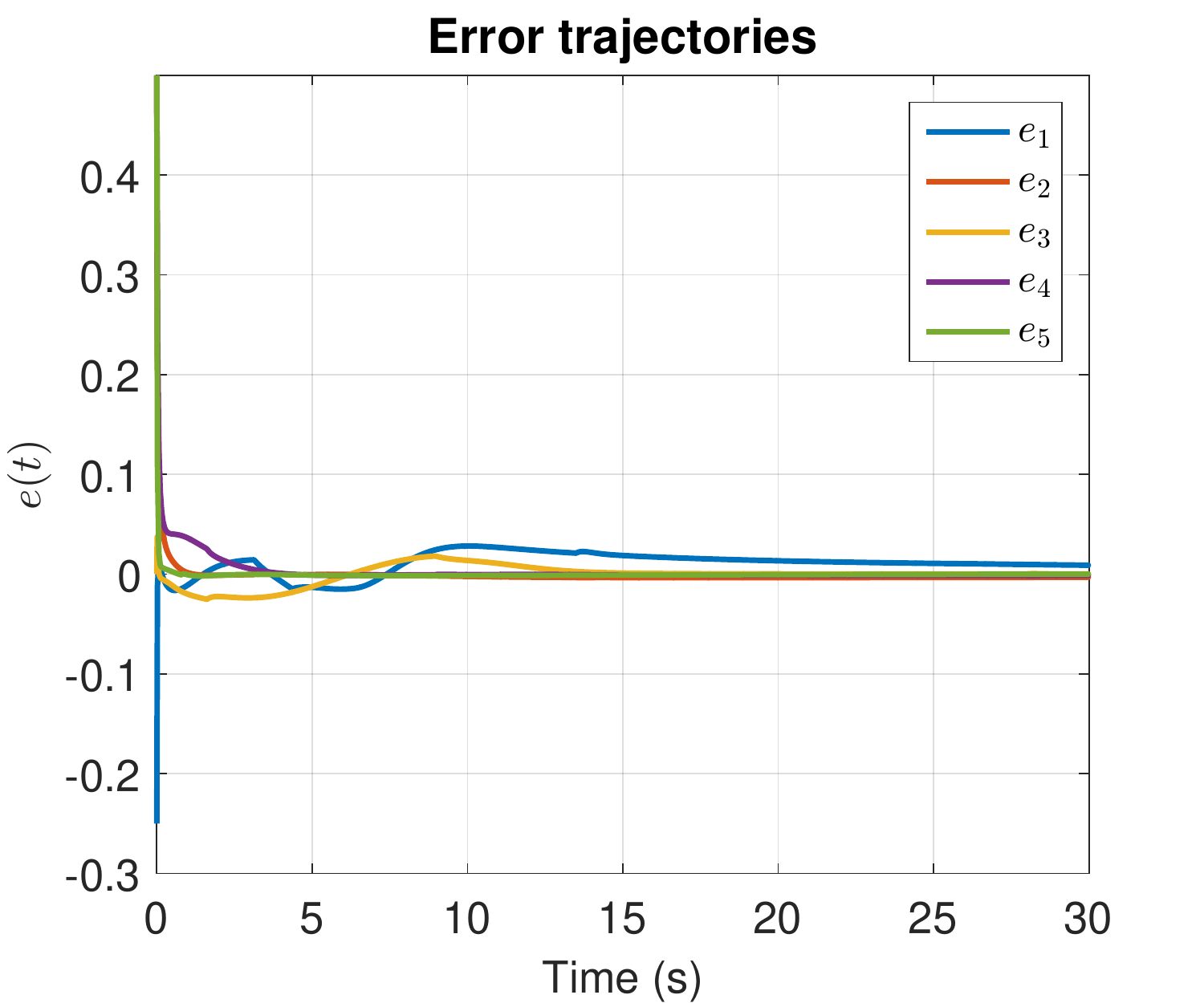} 
\par\end{centering}
\caption{Tracking error trajectories for the agents for the one-dimensional
example.\label{fig:CLNNE}}
\end{figure}
\begin{figure}
\begin{centering}
\includegraphics[width=0.5\columnwidth]{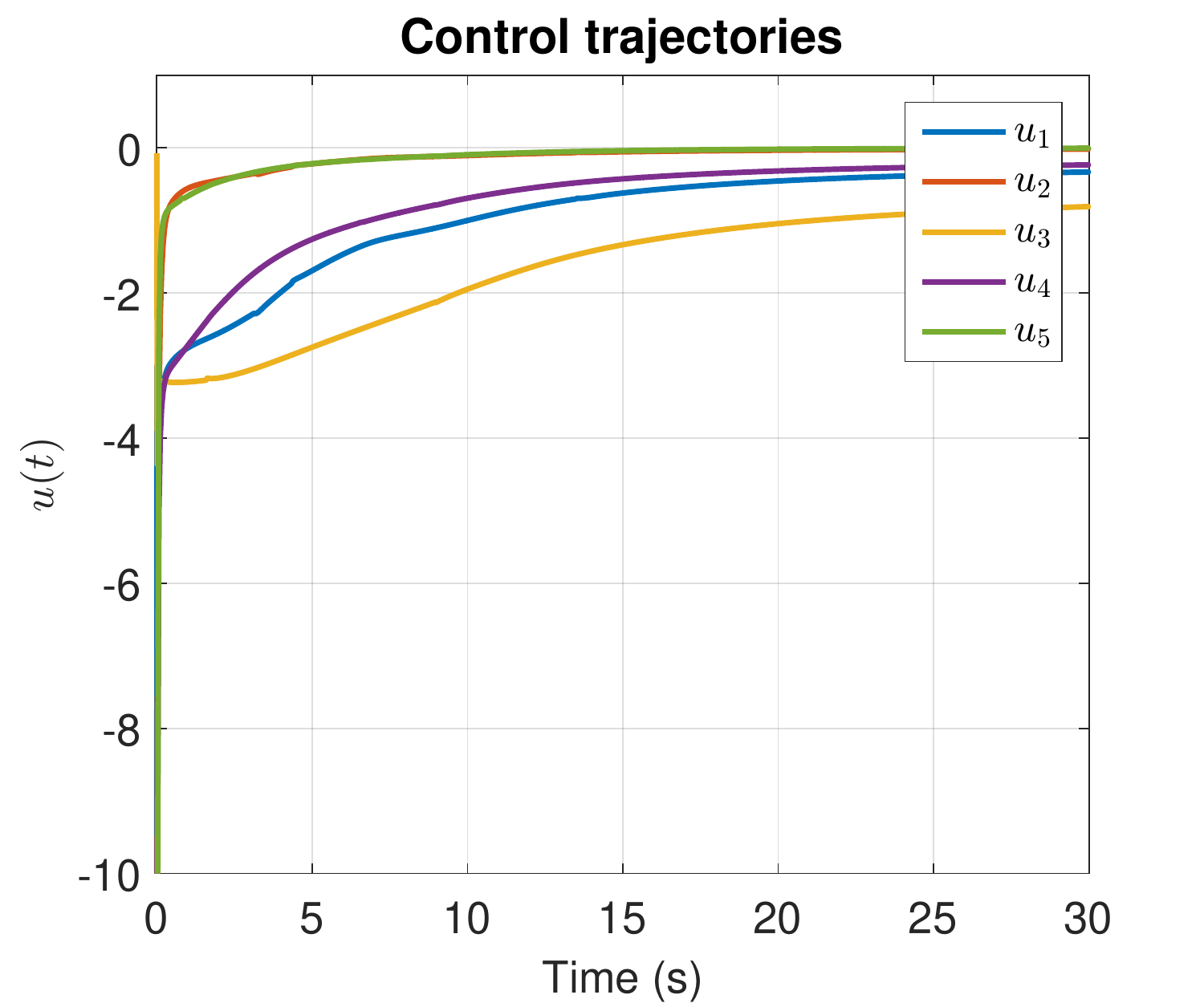}\includegraphics[width=0.5\columnwidth]{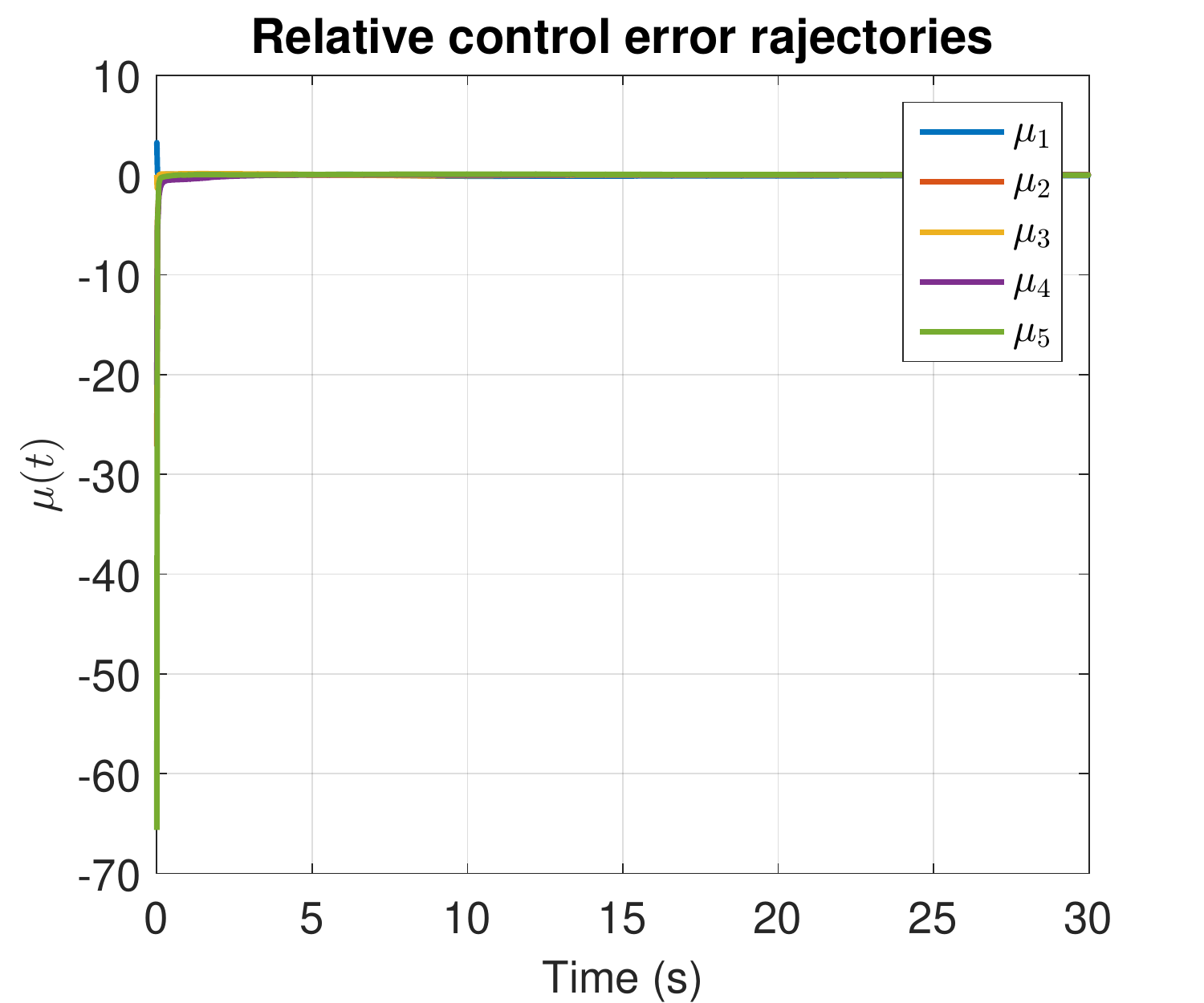} 
\par\end{centering}
\caption{\label{fig:CLNNUandMU}Trajectories of the control input and the relative
control error for all agents for the one-dimensional example.}
\end{figure}
\begin{figure}
\begin{centering}
\includegraphics[width=0.5\columnwidth]{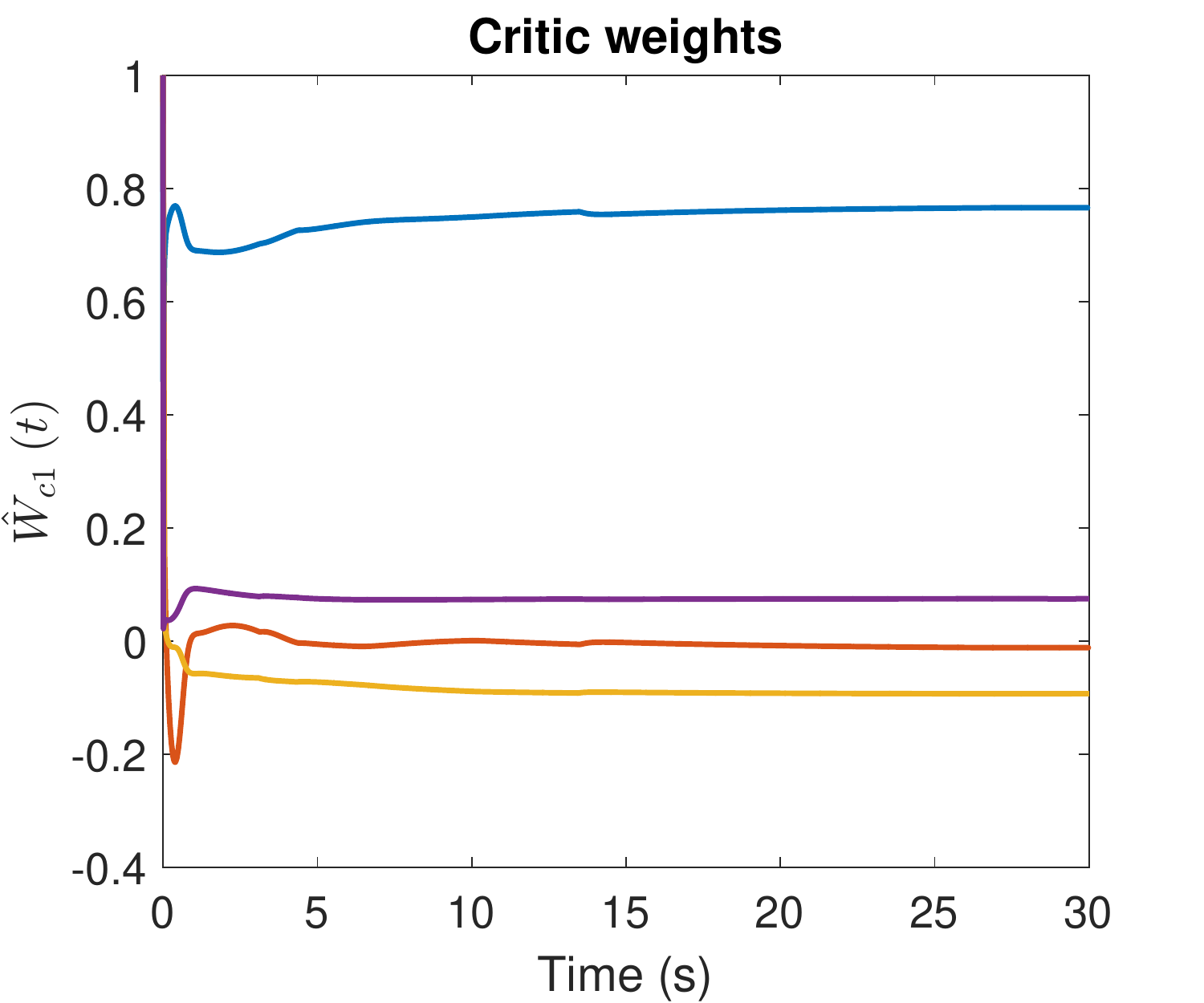}\includegraphics[width=0.5\columnwidth]{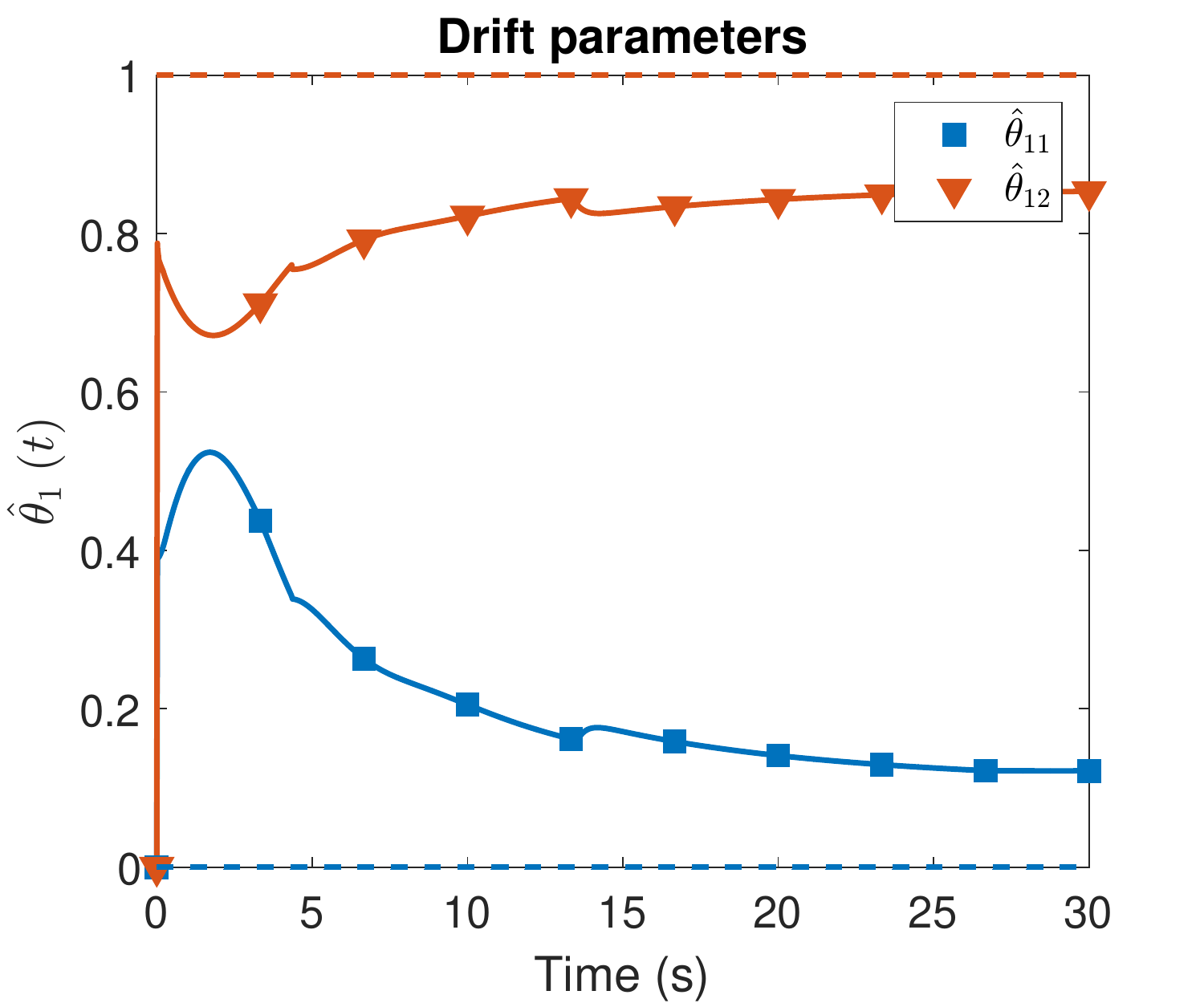} 
\par\end{centering}
\caption{\label{fig:CLNNWCTH1}Value function weights and drift dynamics parameters
estimates for Agent 1 for the one-dimensional example. The dotted
lines in the drift parameter plot are the ideal values of the drift
parameters.}
\end{figure}
Figures \ref{fig:CLNNX} - \ref{fig:CLNNUandMU} show the tracking
error, the state trajectories compared with the desired trajectories,
and the control inputs for all the agents demonstrating convergence
to the desired formation and the desired trajectory. Note that Agents
2, 4, and 5 do not have a communication link to the leader, nor do
they know their desired relative position with respect to the leader.
The convergence to the desired formation is achieved via cooperative
control based on decentralized objectives. Figure \ref{fig:CLNNWCTH1}
shows the evolution and convergence of the value function weights
and the parameters estimates for the drift dynamics for Agent 1. The
errors between the ideal drift parameters and their respective estimates
are large, however, as demonstrated by Figure \ref{fig:CLNNE}, the
resulting dynamics are sufficiently close to the actual dynamics for
the developed technique to generate stabilizing policies. It is unclear
whether the value function and the policy weights converge to their
ideal values. Since an alternative method to solve this problem is
not available to the best of the author's knowledge, a comparison
between value function and policy weight estimates and their corresponding
ideal values is infeasible. %

\section{Concluding Remarks}

A simulation-based actor-critic-identifier architecture is developed
to obtain feedback-Nash equilibrium solutions to a class of differential
graphical games\textit{. }It is established that in a cooperative
game based on minimization of the local neighborhood tracking errors,
the value function corresponding to an agent depends on information
obtained from all their extended neighbors. A set of coupled HJ equations
are developed that serve as necessary and sufficient conditions for
feedback-Nash equilibrium, and closed-form expressions for the feedback-Nash
equilibrium policies are developed based on the HJ equations. The
fact that the developed technique requires each agent to communicate
with all of its extended neighbors motivates the search for a decentralized
method to generate feedback-Nash equilibrium policies.

\begin{IEEEbiography}[{\includegraphics[width=1\columnwidth]{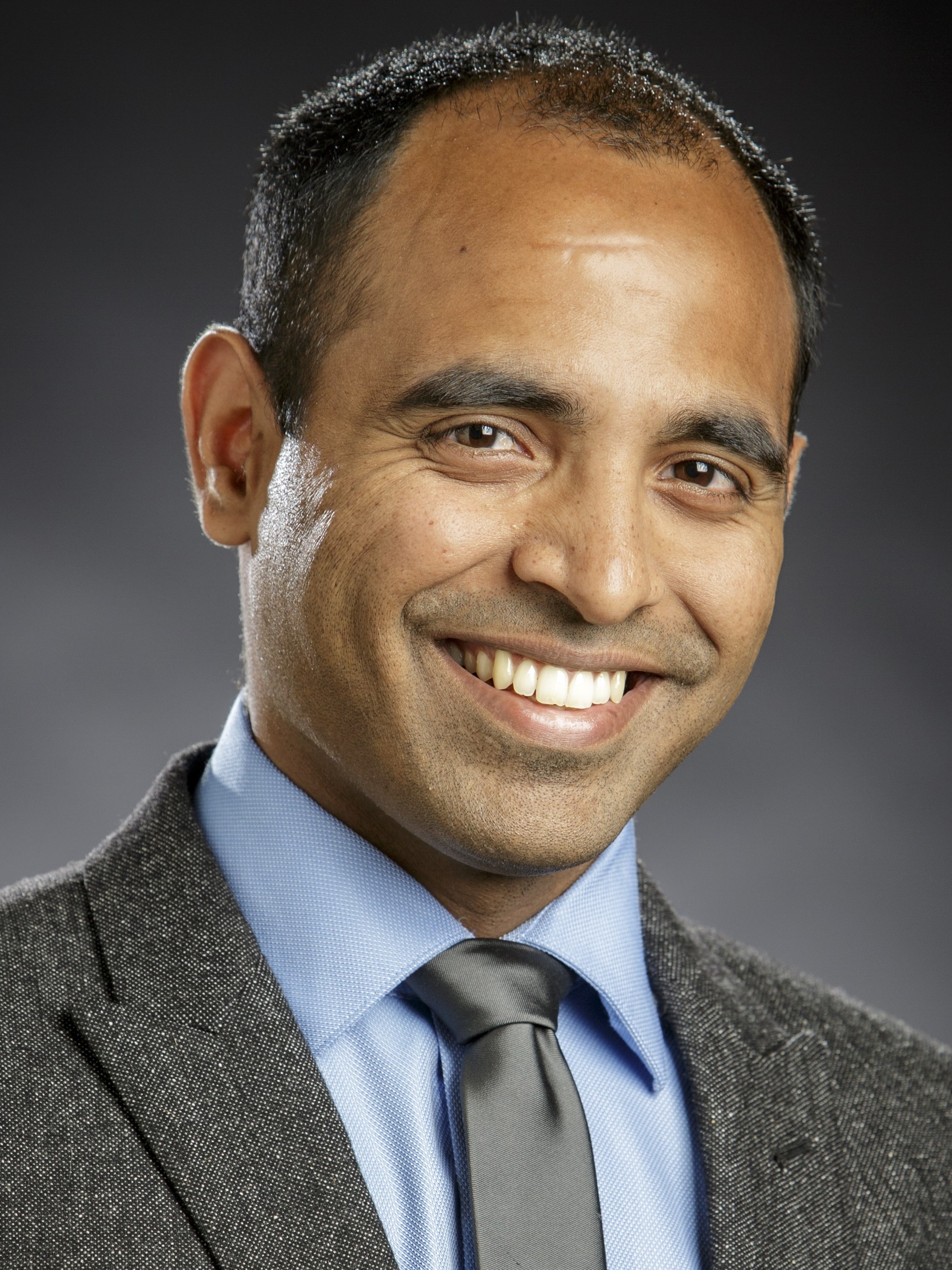}}]{Rushikesh Kamalapurkar}
 received his M.S. and his Ph.D. degree in 2011 and 2014, respectively,
from the Mechanical and Aerospace Engineering Department at the University
of Florida. After working for a year as a postdoctoral research fellow
with Dr. Warren E. Dixon, he was selected as the 2015-16 MAE postdoctoral
teaching fellow. In 2016 he joined the School of Mechanical and Aerospace
Engineering at the Oklahoma State University as an Assistant professor.
His primary research interest has been intelligent, learning-based
control of uncertain nonlinear dynamical systems. His work has been
recognized by the 2015 University of Florida Department of Mechanical
and Aerospace Engineering Best Dissertation Award, and the 2014 University
of Florida Department of Mechanical and Aerospace Engineering Outstanding
Graduate Research Award.
\end{IEEEbiography}

\begin{IEEEbiography}[{\includegraphics[width=1\columnwidth]{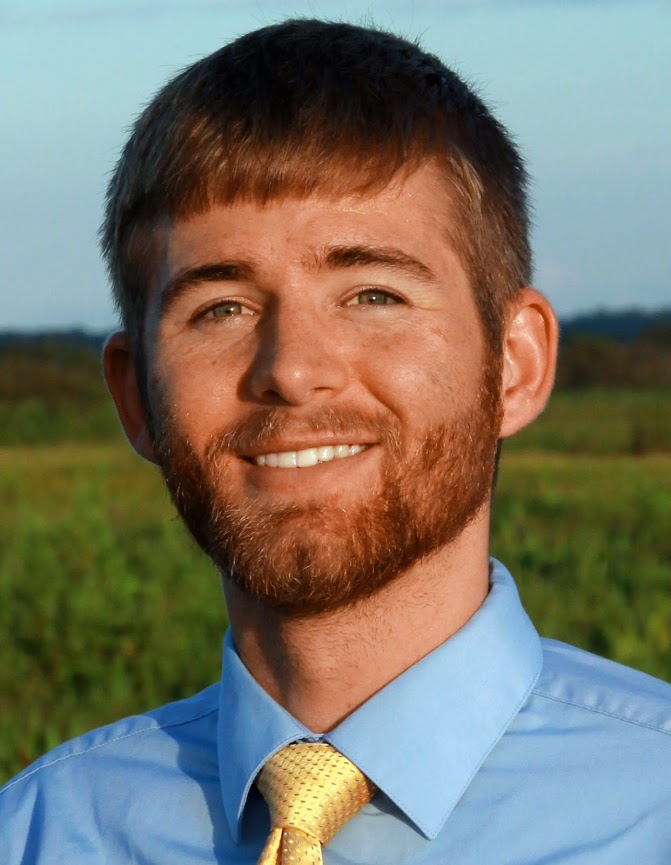}}]{Justin R. Klotz}
 received the Ph.D. degree in mechanical engineering from the University
of Florida, Gainesville, FL, USA, in 2015, where he was awarded the
Science, Mathematics and Research for Transformation (SMART) Scholarship,
sponsored by the Department of Defense. His research interests include
the development of Lyapunov-based techniques for reinforcement learning-based
control, switching control methods, delay-affected control, and trust-based
cooperative control.
\end{IEEEbiography}

\begin{IEEEbiography}[{\includegraphics[width=1\columnwidth]{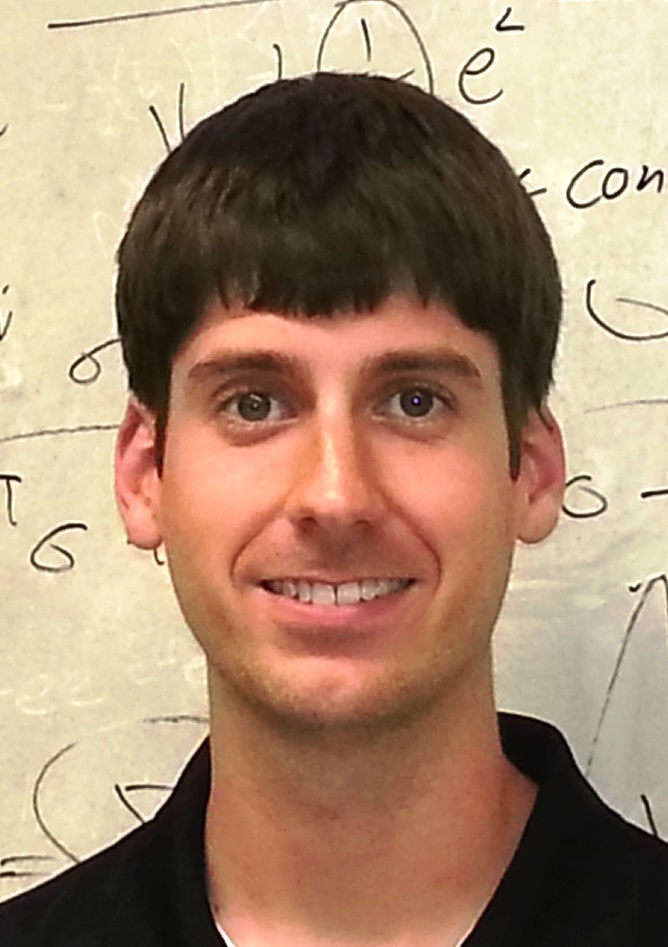}}]{Patrick Walters}
 received the Ph.D. degree in mechanical engineering from the University
of Florida, Gainesville, FL, USA, in 2015. His research interests
include reinforcement learning-based feedback control, approximate
dynamic programming, and robust control of uncertain nonlinear systems
with a focus on the application of underwater vehicles. 
\end{IEEEbiography}

\begin{IEEEbiography}[{\includegraphics[width=1\columnwidth]{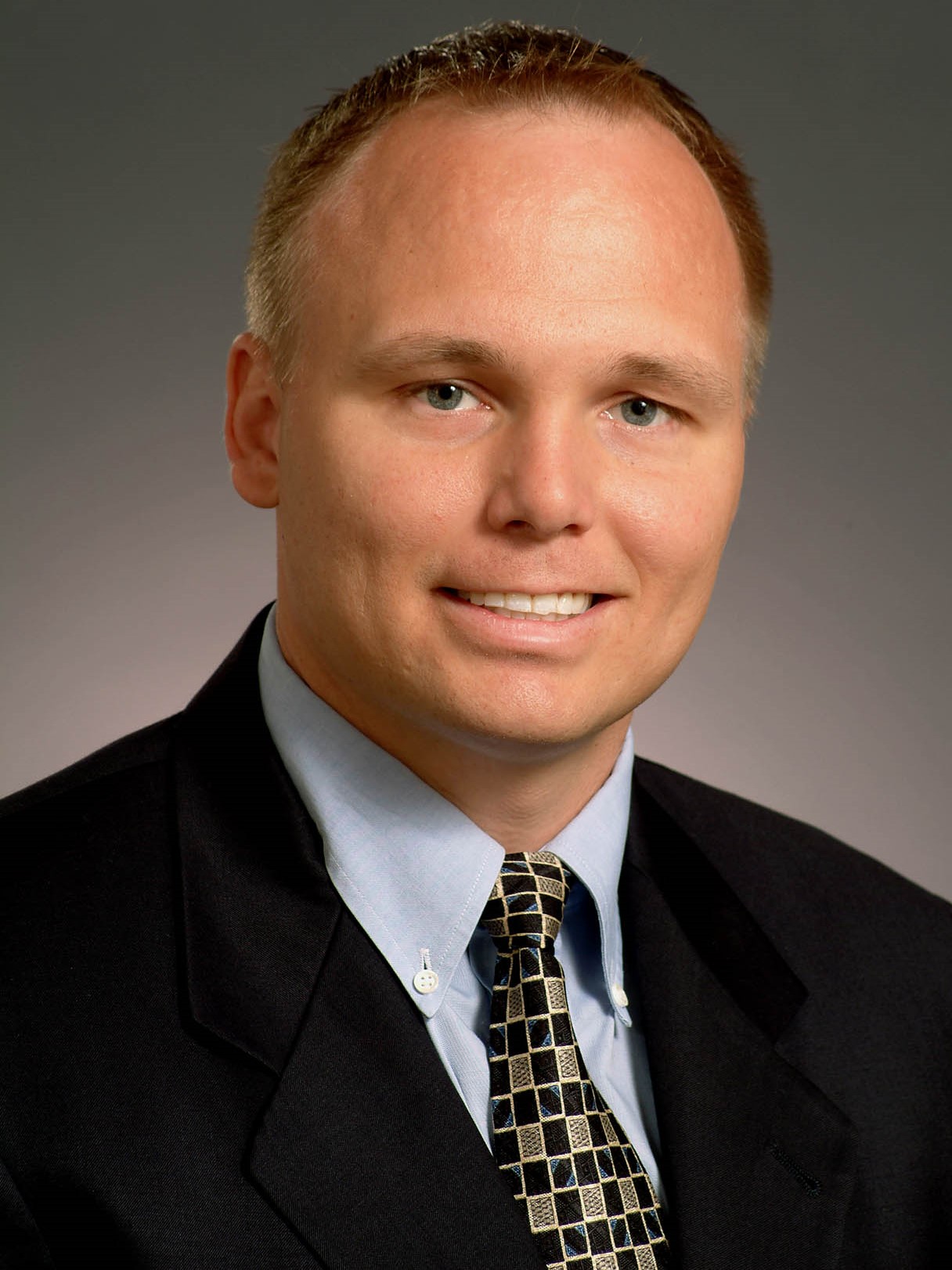}}]{Prof. Warren E. Dixon}
 received his Ph.D. in 2000 from the Department of Electrical and
Computer Engineering from Clemson University. He was selected as a
Eugene P. Wigner Fellow at Oak Ridge National Laboratory (ORNL). In
2004, he joined the University of Florida in the Mechanical and Aerospace
Engineering Department. His main research interest has been the development
and application of Lyapunov-based control techniques for uncertain
nonlinear systems. He has published 3 books, over a dozen chapters,
and approximately 125 journal and 230 conference papers. His work
has been recognized by the 2015 \& 2009 American Automatic Control
Council (AACC) O. Hugo Schuck (Best Paper) Award, the 2013 Fred Ellersick
Award for Best Overall MILCOM Paper, a 2012-2013 University of Florida
College of Engineering Doctoral Dissertation Mentoring Award, the
2011 American Society of Mechanical Engineers (ASME) Dynamics Systems
and Control Division Outstanding Young Investigator Award, the 2006
IEEE Robotics and Automation Society (RAS) Early Academic Career Award,
an NSF CAREER Award, the 2004 Department of Energy Outstanding Mentor
Award, and the 2001 ORNL Early Career Award for Engineering Achievement.
He is a Fellow of ASME and IEEE and is an IEEE Control Systems Society
(CSS) Distinguished Lecturer. He has served as the Director of Operations
for the Executive Committee of the IEEE CSS Board of Governors and
as a member of the U.S. Air Force Science Advisory Board. He is currently
or formerly an associate editor for ASME Journal of Journal of Dynamic
Systems, Measurement and Control, Automatica, IEEE Control Systems
Magazine, IEEE Transactions on Systems Man and Cybernetics: Part B
Cybernetics, and the International Journal of Robust and Nonlinear
Control. 
\end{IEEEbiography}

\end{document}